%
\input harvmac.tex
\noblackbox
%
%
%
%
\def\CC{{\cal C}}

\def\CH{{\cal H}}
\def\CK{{\cal K}}
\def\CL{{\cal L}}
\def\CM{{\cal M}}
\def\CN{{\cal N}}
\def\CS{{\cal S}}
\def\CP{{\cal P}}

\def\CW{{\cal W}}
\def\SLZ{{$SL(2,\IZ)$}}

\def\IZ{\relax\ifmmode\mathchoice
{\hbox{\cmss Z\kern-.4em Z}}{\hbox{\cmss Z\kern-.4em Z}}
{\lower.9pt\hbox{\cmsss Z\kern-.4em Z}}
{\lower1.2pt\hbox{\cmsss Z\kern-.4em Z}}\else{\cmss Z\kern-.4em
Z}\fi}
\def\IB{\relax{\rm I\kern-.18em B}}
\def\IC{{\relax\hbox{$\inbar\kern-.3em{\rm C}$}}}
\def\ID{\relax{\rm I\kern-.18em D}}
\def\IE{\relax{\rm I\kern-.18em E}}
\def\IF{\relax{\rm I\kern-.18em F}}
\def\IG{\relax\hbox{$\inbar\kern-.3em{\rm G}$}}
\def\IGa{\relax\hbox{${\rm I}\kern-.18em\Gamma$}}
\def\IH{\relax{\rm I\kern-.18em H}}
\def\II{\relax{\rm I\kern-.18em I}}
\def\IK{\relax{\rm I\kern-.18em K}}
\def\IP{\relax{\rm I\kern-.18em P}}

\def\inbar{\,\vrule height1.5ex width.4pt depth0pt}
\def\p{\partial}
\def\pb{{\bar \p}}

\font\cmss=cmss10 \font\cmsss=cmss10 at 7pt
\def\IR{\relax{\rm I\kern-.18em R}}

\def\liet{{\underline{\bf t}}}

\def\sgn{{\rm sgn}}
\def\sdtimes{\mathbin{\hbox{\hskip2pt\vrule
height 4.1pt depth -.3pt width .25pt\hskip-2pt$\times$}}}
\def\Tr{\rm Tr}

\def\wb{{\bar{w}}}

\def\ymt{$YM_2$}
\def\zb {{\bar{z}}}
\def\CW{\cal W}
\def\CZ{{\cal Z}}
\def\imysq{(\Im y)^2}
\def\imyvsq{(\Im \vec y)^2}
\def\ymt{y_{-,2}}
\def\ypt{y_{+,2}}
\def\yv{\vec{y}}
\def\yvb{\vec{y}^*}
\def\hone{h^{(1)} }
%
%

\lref\blst{D. Bailin, A. Love, W. A. Sabra and S. Thomas, 
``Duality symmetries of threshold corrections in orbifold models,''
Phys. Lett. {\bf B320} (1994) 21 \semi ``String loop threshold
corrections for Z(N) Coxeter orbifolds,'' Mod. Phys. Lett. {\bf A9}
(1994) 67. }

\lref\antoni{I. Antoniadis, S. Ferrara, E. Gava, K.S. Narain and T.R. Taylor,
``Perturbative Prepotential and Monodromies in N=2 Heterotic Superstring,''
Nucl. Phys. {\bf B447} (1995) 35, hep-th/9504034.}

\lref\antonii{ I. Antoniadis and T.R. Taylor,
``String loop corrections to gauge and Yukawa couplings,''
hep-th/9301033.}

\lref\bksdx{T. Banks and L. Dixon, ``Constraints on string vacua with
space-time supersymmetry,'' Nucl. Phys. {\bf B307} (1988) 93. }

\lref\borcha{R. E. Borcherds, ``The monster Lie algebra,'' Adv. Math. {\bf 83}
No. 1 (1990).}

\lref\borchi{R. Borcherds,``Monstrous moonshine
and monstrous Lie superalgebras,'' Invent. Math.
{\bf 109}(1992) 405.}

\lref\borchii{R. Borcherds, ``Automorphic forms
on $O_{s+2,2}(R)$ and infinite products,''
Invent. Math. {\bf 120}(1995) 161.}

\lref\borchiii{R. Borcherds,
``Automorphic forms
on $O_{s+2,2}(R)^+$ and generalized Kac-Moody
algebras,'' contribution to the Proceedings of
the 1994 ICM, Zurich. }

\lref\borchiv{R. Borcherds, ``The moduli space
of Enriques surfaces and the fake monster Lie
superalgebra,''  preprint (1994).}

\lref\borchalg{R. Borcherds, ``Generalized Kac-Moody algebras,'' Journal of
Algebra
{\bf 115} (1988) 501.}

\lref\clm{G. L. Cardoso, D. L\"{u}st and T. Mohaupt,
``Threshold corrections and symmetry enhancement in string
compactifications,'' Nucl. Phys. {\bf B450} (1995) 115,
hep-th/9412209.}

\lref\dewit{B. de Wit, V. Kaplunovsky, J. Louis and D.  L\"{u}st,
``Perturbative Couplings of Vector Multiplets in $N=2$ Heterotic String
Vacua,''
Nucl. Phys. {\bf B451} (1995) 53, hep-th/9504006.}

\lref\eguchi{T. Eguchi, H. Ooguri, A. Taormina and S. K. Yang, ``Superconformal
algebras and string compactification on manifolds with $SU(n)$ holonomy,''
Nucl. Phys. {\bf B315} (1989) 193. }

\lref\egtaor{T. Eguchi and A. Taormina, ``Character formulas for the $N=4$
superconformal
algebra,'' Phys. Lett. {\bf 200B} (1988) 315. }

\lref\fklz{S. Ferrara, C. Kounnas, D. L\"{u}st and F. Zwirner,
``Duality-invariant
partition functions and automorphic superpotentials for $(2,2)$ string
compactifications,''  Nucl. Phys.  {\bf B365} (1991) 431. }

\lref\mayrst{P. Mayr and S. Stieberger, ``Threshold corrections to
gauge couplings in orbifold compactifications,''  Nucl. Phys. {\bf B407} (1993)
725,
hep-th/9303017. }

\lref\mayrsti{P. Mayr and S. Steiberger, ``Moduli dependence of one loop gauge
couplings in $(0,2)$ compactifications,'' Phys. Lett. {\bf B355} (1995) 107,
hep-th/9504129.}

\lref\oogv{H. Ooguri and C. Vafa, ``Geometry of $N=2$ strings,''
Nucl. Phys. {\bf B361} (1991) 969.}

\lref\oogvi{H. Ooguri and C. Vafa,  `` $N=2$ heterotic strings, '' Nucl. Phys.
{\bf B367} (1991) 83.}

\lref\ferrarai{ A.\ Ceresole, R.\ D'Auria, S.\ Ferrara and A.\ Van Proeyen,
``On Electromagnetic Duality in Locally Supersymmetric N=2 Yang--Mills
Theory,''
hep-th/9412200.}

\lref\fhsv{
S. Ferrara, J. A. Harvey, A. Strominger, C. Vafa ,
``Second-Quantized Mirror Symmetry, ''
hep-th/9505162. }

\lref\vaftest{C. Vafa, ``A stringy test of the fate of the conifold,''  Nucl.
Phys.
{\bf B447} (1995) 252, hep-th/9505053.}

\lref\gebert{R.W. Gebert,
``Introduction to Vertex Algebras, Borcherds Algebras, and the Monster Lie
Algebra,''
Int. J. Mod. Phys. {\bf A8} (1993) 5441, hep-th/9308151.}

\lref\gnw{R.W. Gebert, H. Nicolai and P.C. West,
``Multistring Vertices and Hyperbolic Kac Moody Algebras,''
hep-th/9505106.}

\lref\jorgenson{J. Jorgenson and A. Todorov,
``A conjectured analog of Dedekind's eta function
for K3 surfaces,'' Yale preprint. }

\lref\kv{S. Kachru and C. Vafa, ``Exact results for $N=2$ compactifications of
heterotic strings, '' Nucl. Phys. {\bf B450} (1995) 69, hep-th/9505105. }

\lref\lercheeg{W. Lerche, ``Elliptic index and superstring effective
actions,'' Nucl. Phys. {\bf B308} (1988) 102. }

\lref\lsw{see e.g. W. Lerche, A. N. Schellekens and N. P. Warner,
`` Lattices and Strings, '' Phys. Rep. {\bf 177}  (1989) 1. }

\lref\schwarn{A. N. Schellekens and N. P. Warner, Phys. Lett. {\bf 177B}
(1986) 317; Phys. Lett. {\bf 181B} (1986) 339; Nucl. Phys. {\bf B287} (1987)
317.}

\lref\moorei{ G. Moore,
``Finite in All Directions, '' hep-th/9305139; G.  Moore, ``Symmetries and
symmetry-breaking in string theory,'' hep-th/9308052.}

\lref\mooreii{G. Moore, ``Symmetries of the Bosonic String S-Matrix,''
hep-th/9310026; Addendum to: ``Symmetries of the Bosonic String S-Matrix,''
hep-th/9404025.}

\lref\nikulini{
V. A. Gritsenko, V. V. Nikulin,
``Siegel automorphic form corrections of some Lorentzian Kac--Moody Lie
algebras, ''
alg-geom/9504006.}

\lref\nikulinii{V. V. Nikulin,
``Reflection groups in hyperbolic spaces and the
denominator formula for Lorentzian Kac--Moody Lie algebras,''
alg-geom/9503003.}

\lref\nikuliniii{
V. A. Gritsenko, V. V. Nikulin, ``K3 Surfaces,
Lorentzian Kac-Moody Algebras, and
Mirror Symmetry,'' alg-geom/9510008.}

\lref\witteg{E. Witten, ``Elliptic Genera and Quantum
Field Theory,'' Commun. Math. Phys. {\bf 109}(1987)525;
``The index of the Dirac operator in loop space,''  Proceedings of the
conference
on elliptic curves and modular forms in algebraic topology, Princeton NJ,
1986.}

\lref\swa{N. Seiberg and E. Witten, ``Electric-magnetic duality, monopole
condensation, and
confinement in $N=2$ supersymmetric Yang-Mills theory,'' Nucl. Phys. {\bf B426}
(1994) 19; (E) {\bf B340} (1994) 485, hep-th/9407087. }

\lref\swb{N. Seiberg and E. Witten, `` Monopoles, duality and chiral symmetry
breaking in $N=2$ supersymmetric QCD, '' Nucl. Phys. {\bf B431} (1994) 484,
hep-th/9408099. }

\lref\dabh{A. Dabholkar and J. A. Harvey, ``Nonrenormalization of the
superstring
tension,'' Phys. Rev. Lett. {\bf 63} (1989) 478; A. Dabholkar, G. Gibbons,
J. A. Harvey and F. Ruiz Ruiz, ``Superstrings and solitons,''
Nucl. Phys. {\bf B340} (1990) 33.}

\lref\sschwarz{J. Schwarz and A. Sen, ``Duality symmetries of 4-D heterotic
strings,''
Phys. Lett. {\bf B312} (1993) 105, hep-th/9305185.}

\lref\hullt{C. Hull and P. Townsend, ``Unity of superstring dualities,'' Nucl.
Phys.
{\bf B438} (1995) 109, hep-th/9410167.}

\lref\wittdyn{E. Witten, ``String theory dynamics in various dimensions,''
Nucl.
Phys. {\bf B443} (1995) 85, hep-th/9503124.}

\lref\klti{A. Klemm, W. Lerche and S. Theisen, ``Nonperturbative effective
actions of
$N=2$ supersymmetric gauge theories,'' hepth-9505150. }

\lref\nsi{S. Cecotti, P. Fendley, K. Intriligator and C. Vafa, ``A new
supersymmetric
index, '' Nucl. Phys. {\bf B386} (1992) 405, hep-th/9204102;
S. Cecotti and C. Vafa, ``Ising model
and $N=2$ supersymmetric theories, '' Commun.  Math. Phys. {\bf 157} (1993)
139,
hep-th/9209085.}

\lref\walton{M.  A. Walton,  ``Heterotic string on the simplest Calabi-Yau
manifold and
its orbifold limit, '' Phys. Rev. {\bf D37} (1988) 377. }

\lref\dkl{L. Dixon,  V. S. Kaplunovsky and J. Louis, ``Moduli-dependence of
string
loop corrections to gauge coupling constants, ''
Nucl. Phys. {\bf B329} (1990) 27. }

\lref\vadim{V. Kaplunovsky, ``One loop threshold effects in
string unification,'' Nucl. Phys. {\bf B307} (1988) 145,  revised
in hep-th/9205070.}

\lref\kaplouis{V. Kaplunovsky and J. Louis, ``On gauge couplings in string
theory,''
Nucl. Phys. {\bf B444} (1995) 191, hep-th/9502077. }

\lref\cv{E. Calabi and E. Vesentini, Ann. Math. {\bf 71} (1960) 472.}

\lref\FMS{D. Friedan, E. Martinec,  and S. Shenker,
``Conformal Invariance, Supersymmetry, and String Theory,''
Nucl.Phys. {\bf B271} (1986) 93.}

\lref\gilmore{R. Gilmore, {\it Lie Groups, Lie Algebras and some of their
Applications, }Wiley-Interscience, New York, 1974.}

\lref\patera{ J. Patera, R. T. Sharp and P. Winternitz, J. Math. Phys. {\bf 17}
(1976) 1972. }

\lref\agn{I. Antoniadis,  E. Gava, K.S. Narain, ``Moduli corrections to
gravitational
couplings from string loops,'' Phys. Lett. {\bf B283} (1992) 209,
hep-th/9203071; `` Moduli corrections to gauge and gravitational couplings in
four-dimensional superstrings,'' Nucl. Phys. {\bf B383} (1992) 109,
hep-th/9204030.}

\lref\agnt{I. Antoniadis,
 E. Gava, K.S. Narain and T.R. Taylor,
``Superstring threshhold corrections to
Yukawa couplings,'' Nucl. Phys {\bf B407} (1993) 706;
hep-th/9212045. Note: These versions are
different. }

\lref\kennati{See K. Intriligator and N. Seiberg, ``Lectures on supersymmetric
gauge
theories and electric-magnetic duality,'' hep-th/9509066 for a recent review.}

\lref\fvanp{S. Ferrara and A. Van Proeyen, `` A theorem on $N=2$ special
Kahler product manifolds,'' Class. Quantum Grav. {\bf 6}
(1989) L243. }

\lref\recent{Recent test of duality refs}

\lref\yoshii{H. Yoshii, ``On moduli space of $c=0$ topological conformal
field theories,'' Phys. Lett. {\bf B275} (1992) 70.}

\lref\nojiri{S. Nojiri, ``$N=2$ superconformal topological field theory,''
 Phys. Lett. {\bf 264B}(1991)57. }

\lref\berkvaf{N. Berkovits and C. Vafa, ``$N=4$
topological strings, ''  Nucl. Phys. {\bf B433} (1995) 123, hep-th/9407190.}

\lref\vwdual{ C. Vafa and E. Witten, ``Dual string pairs with $N=1$ and $N=2$
supersymmetry in four dimensions, '' hep-th/9507050. }

\lref\monstref{B. H. Lian and S. T. Yau, ``Arithmetic properties of mirror map
and quantum
coupling, '' hep-th/9411234. }

\lref\lianyau{B. H. Lian and S. T. Yau, ``Arithmetic properties of mirror map
and quantum
coupling, '' hep-th/9411234;
``Mirror Maps, Modular Relations
and Hypergeometric Series I ,'' hep-th/9506210;
``Mirror Maps, Modular
Relations and Hypergeometric Series II,''
hep-th/9507153}

\lref\bcov{M. Bershadsky, S. Cecotti, H. Ooguri and C. Vafa, `` Kodaira-Spencer
theory
of gravity and exact results for quantum string amplitudes, '' Commun. Math.
Phys.
{\bf 165} (1994) 311, hep-th/9309140. }

\lref\fgz{I. Frenkel, H. Garland, and G. Zuckerman, ``Semi-infinite
cohomology and string theory,'' Proc. Nat. Acad. Sci.
{\bf 83}(1986) 8442.}

\lref\gross{ D. Gross, ``High energy symmetries of
string theory,'' Phys. Rev. Lett. {\bf 60B} (1988) 1229.}

\lref\horava{P. Horava, ``Strings on world sheet orbifolds,''  Nucl. Phys.
{\bf B327} (1989) 461; ``Background duality of open string models,'' Phys.
Lett.
{\bf B231}(1989)251.}

\lref\sgntti{M. Bianchi, G. Pradisi and A. Sagnotti, ``Toroidal
compactification and symmetry breaking in open string theories,''
 Nucl. Phys. {\bf B376} (1992) 365.}

\lref\polch{J. Polchinski, ``Combinatorics of boundaries in
string theory,'' Phys. Rev. {\bf D50} (1994) 6041,
hep-th/9407031;  M. B. Green,
``A gas of D instantons,'' Phys. Lett. {\bf B354} (1995) 271,
hep-th/9504108.}

\lref\shenker{S. Shenker, ``Another Length Scale in
String Theory,'' hep-th/9509132}

\lref\argf{P. C. Argyres and A. E.  Faraggi, ``The vacuum structure and
spectrum of
$N=2$ supersymmetric $SU(n)$ gauge theory, '' Phys. Rev. Lett.
{\bf 74} (1995) 3931, hep-th/9411057.}

\lref\givpor{A. Giveon and M. Porrati, ``Duality invariant string algebra and
$D=4$ effective actions, '' Nucl. Phys. {\bf B355} (1991) 422. }

\lref\dfkz{J. P. Derendinger, S. Ferrara, C. Kounnas and F. Zwirner, ``On loop
corrections to string effective field theories: field-dependent gauge couplings
and
$\sigma$-model anomalies,'' Nucl. Phys. {\bf B372} (1992) 145. }

\lref\agnti{I. Antoniadis, E. Gava, K. S. Narain and T. R. Taylor, ``$N=2$ Type
II-
Heterotic duality and higher derivative F-terms, '' hep-th/9507115.}

\lref\klt{V. Kaplunovsky, J. Louis and S. Theisen, ``Aspects of duality in
$N=2$ string vacua,'' Phys. Lett. {\bf B357} (1995) 71, hep-th/9506110.}

\lref\louispas{J. Louis, PASCOS proceedings, P. Nath ed., World
Scientific 1991.}

\lref\lco{G. L. Cardoso and B. A. Ovrut, ``A Green-Schwarz mechanism
for $D=4$, $N=1$ supergravity anomalies,'' Nucl. Phys. {\bf B369} (1992) 351;
``Coordinate and Kahler sigma model anomalies and their cancellation
in string effective field theories,''  Nucl. Phys. {\bf B392} (1993) 315,
hep-th/9205009.}

\lref\levin{L. Lewin, Polylogarithms and Associated Functions,
North Holland 1981. See eq.  6.7.}

\lref\witthyper{E. Witten, ``Topological tools in ten dimensional physics, ''
in
{\it Unified String Theories}, eds. M. Green and D. Gross, World Scientific,
Singapore,
1986.}

\lref\fein{A. Feingold and I. Frenkel,
``A Hyperbolic Kac-Moody Algebra and the
Theory of Siegel Modular Forms of Genus
2,'' Math. Ann. {\bf 263} (1083) 87.}

\lref\frenk{I. Frenkel, ``Representations of Kac-Moody algebras
and dual resonance models,'' in {\it Applications
of Group Theory in Physics and Mathematical Physics},
Vol. 21, Lectures in Applied Mathematics, M. Flato,
P. Sally, G. Zuckerman, eds. AMS 1985.}

\lref\golatt{P. Goddard and D. Olive, ``Algebras, Lattices,
and Strings,''
in {\it Vertex operators in mathematics and
physics},'' ed. J. Lepowsky et. al. Springer-Verlag, 1985.}

\lref\gebnic{R. W. Gebert and H. Nicolai, `` On $E_{10}$ and the DDF
construction,''
hep-th/9406175.}

\lref\wittorb{E. Witten, ``Space-time and topological orbifolds,'' Phys. Rev.
Lett.
{\bf 61} (1988) 670.}

\lref\fuchs{J. Fuchs, {\it Affine lie algebras and quantum groups,} Cambridge
University Press,  Cambridge, 1992. }

\lref\kac{V. G. Kac, {\it Infinite dimensional Lie algebras,} Cambridge
University Press, Cambridge, 1990. }

\lref\moonshine{J. H. Conway and S. P. Norton, ``Monstrous moonshine,''
Bull. London Math. Soc. {\bf 11} (1979) 308. }

\lref\nikulin{see e.g. V. V. Nikulin, ``Reflection groups in hyperbolic spaces
and the denominator formula for Lorentzian Kac-Moody algebras,
alg-geom/9503003.}

\lref\kmw{V. G. Kac, R. V. Moody, and M. Wakimoto, ``On $E_{10}$,'' In K.
Bleuler
and M. wener, eds. {\it Differential geometrical methods in theoretical
physics.}
Proceedings, NATO advanced research workshop, 16th international conference,
Como
Kluwer, 1988.}

\lref\gebnici{R. W. Gebert and H. Nicolai, ``$E_{10}$ for beginners,''
hep-th/9411188.}

\lref\flm{I. B. Frenkel, J. Lepowsky, and A. Meurman, {\it Vertex operator
algebras
and the monster,} Pure and Applied Mathematics Volume 134, Academic
Press, San Diego, 1988.}

\lref\lianzuck{B. H. Lian and  G. J. Zuckerman, ``New
Perspectives on the BRST Algebraic Structure of
String Theory,''  Commun.Math.Phys. {\bf 154} (1993) 613,
hep-th/9211072.}

\lref\mool{C. Montonen and D. Olive, ``Magnetic monopoles as
gauge particles? ''Phys. Lett. {\bf 72B} (1977)
117; P. Goddard, J. Nuyts and D. Olive, ``Gauge theories
and magnetic charge,''  Nucl. Phys. {\bf B125} (1977) 1.}

\lref\WO{E. Witten and D. Olive, ``Supersymmetry algebras that
include topological charges,'' Phys. Lett. {\bf 78B} (1978) 97. }

\lref\senb{A. Sen, ``Dyon-monopole bound states, selfdual harmonic
forms on the multi-monopole moduli space, and $SL(2,Z)$ invariance
in string theory,'' Phys. Lett. {\bf 329} (1994) 217,
hep-th/9402032.}

\lref\osborn{H. Osborn, ``Topological charges for
$N=4$ supersymmetric gauge theories and monopoles of
spin 1,'' Phys. Lett. {\bf 83B} (1979) 321.}

\lref\andycone{A. Strominger, ``Massless black holes and conifolds in
string theory,'' Nucl. Phys. {\bf B451} (1995) 96; hep-th/9504090.}

\lref\coneheads{B. R. Greene, D. R. Morrison and A. Strominger, ``Black hole
condensation and the unification of string vacua,'' Nucl. Phys. {\bf B451}
(1995) 109; hep-th/9504145}

\lref\joebrane{J. Polchinski, ``Dirichlet-Branes and Ramond-Ramond charges''
hep-th/9510017.}

\lref\sena{For reviews and further references
see M. Duff, R. Khuri and J. X. Lu,
``String Solitons'' Phys. Rep. {\bf 259} (1995) 213, hep-th/9412184;
 A. Sen, ``Strong-weak coupling duality in four-dimensional string
theory,'' Int. J. Mod. Phys. {\bf A9} (1994) 3707.}

\lref\givrev{A. Giveon, M. Porrati and E. Rabinovici, ``Target space duality in
string
theory,'' Phys. Rep. {\bf 244} (1994) 77;hep-th/9401139}

\lref\kirka{E. Kiritsis and C. Kounnas, ``Infrared Regularization of
superstring
theory and the one-loop calculation of coupling constants,'' Nucl. Phys.
{\bf B442} (1995) 442, hep-th/9501020.}

\lref\min{J. Minahan, Nucl. Phys. {\bf B298} (1988) 36.}

\lref\afgntrev{I. Antoniadis, S. Ferrara, E. Gava, K. S. Narain and T. R.
Taylor,
``Duality symmetries in $N=2$ heterotic superstring,'' hep-th/9510079.}

\lref\bjulia{B. Julia, in {\it Applications of group theory in physics
and mathematical physics,} ed. P. Sally et. al. (American Mathematical
Society, Providence, 1985). }

\lref\dwvp{B. de Wit and A. Van Proeyen, ``Broken
sigma-model isometries in very special
geometry,'' Phys. Lett. {\bf 293B}(1992)94.}

\lref\klm{A. Klemm, W. Lerche and P. Mayr, ``K3-fibrations and
Heterotic-Type II string duality, '' Phys. Lett. {\bf B357}
(1995) 313, hep-th/9506122.}

\lref\cogp{P. Candelas, X. de la Ossa, P. S. Green and L. Parkes,
``A pair of Calabi-Yau manifolds as an exactly soluble
superconformal theory,'' Nucl. Phys. {\bf B359} (1991) 21.}

\lref\gvz{M. T. Grisaru, A. van de Ven and D. Zanon, Phys. Lett.
{\bf 173B} (1986) 423, Nucl. Phys. {\bf B277} (1986) 388,
Nucl. Phys. {\bf B277} (1986) 409.}

\lref\witzw{E. Witten and B. Zwiebach, ``Algebraic
structures and differential geometry in 2d string theory,''
Nucl. Phys. {\bf B377}(1992)55;hep-th/9201056}

\lref\cdfkm{P. Candelas, X. de la Ossa, A. Font, S. Katz,
and D. Morrison, ``Mirror Symmetry for Two Parameter
Models - I '' Nucl. Phys. {\bf B416} (1994) 481, hep-th/9308083.}

\lref\iban{G. Aldazabal, A. Font, L. E. Ib\'{a}\~{n}ez and F. Quevedo,
``Chains of $N=2$,$D=4$ heterotic/type II duals,'' hep-th/9510093.}

\lref\wip{Work in progress.}

\lref\martinec{E. Martinec, ``Criticality, Catastrophes, and
Compactifications,'' in L. Brink et. al. eds. {\it Physics and
Mathematics of Strings}, V. Knizhnik memorial volume.
World Scientific, 1990.}

\lref\aspmorr{P.S. Aspinwall and D.R. Morrison,
``String Theory on K3 Surfaces,''  hep-th/9404151.}

%
%

\Title{\vbox{\baselineskip12pt
\hbox{hep-th/9510182}
\hbox{EFI-95-64}
\hbox{YCTP-P16-95}
}}
{\vbox{\centerline{Algebras, BPS States, and Strings } }}

\centerline{Jeffrey A. Harvey}
\bigskip
\centerline{\sl Enrico Fermi Institute, University of Chicago}
\centerline{\sl 5640 Ellis Avenue, Chicago, IL 60637 }
\centerline{\it harvey@poincare.uchicago.edu}
\bigskip
\centerline{Gregory Moore}
\bigskip
\centerline{\sl Department of Physics, Yale University}
\centerline{\sl New Haven, CT  06511}
\centerline{ \it moore@castalia.physics.yale.edu }

\bigskip
\centerline{\bf Abstract}
We  clarify the role played
by BPS states in  the calculation of threshold corrections
of D=4, N=2 heterotic string compactifications. We evaluate
these corrections for some classes of compactifications
and show that they are sums of logarithmic functions over
the positive roots of generalized Kac-Moody algebras.
Moreover, a certain limit of the formulae suggests a
reformulation of heterotic string in terms of a gauge theory
based on hyperbolic algebras such as $E_{10}$.
We define a generalized Kac-Moody Lie superalgebra associated
to the BPS states.  Finally we discuss the relation of our
results with string duality.

\Date{October 24, 1995, Revised Jan. 9, 1996}
%

\newsec{Introduction}

It has become clear recently that theories with  extended supersymmetry
have a rich dynamical structure which is nonetheless amenable to exact analysis
using ideas of electromagnetic duality \refs{\mool, \WO, \osborn, \senb,
\swa,\swb,\argf,\klti}. One important feature of these
theories is the existence of BPS  states. These states play a crucial role in
the
dynamics of the theory and their structure at strong coupling can in many
cases be determined by a semi-classical analysis.
{}For example, in the analysis of \refs{\swa, \swb}
certain BPS states become massless at special points in the quantum moduli
space of vacua and dominate the low-energy dynamics.

In string theory there is increasing evidence that a related although
undoubtedly
richer structure is present, particularly in $N=2$ theories
exhibiting duality \refs{\kv, \fhsv}.
String theories with  extended supersymmetry resulting from toroidal
compactification
have in fact an infinite spectrum of BPS states \refs{\dabh, \sschwarz}.
These BPS states have played a central role in much of the recent
work on duality in string theory.
In Type II string theory the presence of
an infinite tower of non-perturbative BPS states is crucial in
understanding the strong coupling behavior and duality symmetries
\refs{\hullt, \wittdyn}.  These BPS states carrying Ramond-Ramond
charge are also essential to the resolution of the conifold singularity
in Type II string theory \refs{\andycone,\coneheads} and play an
important role in  understanding
the relation between Type I and Type II string theory \joebrane. In the
heterotic
string BPS states played a central role in the original understanding
of $S$-duality \sena.
It seems fair to say that the foundations of string theory are shifting and
that the structure of BPS states provides one of the most useful clues
as to what type of theory unifies these disparate phenomena.

Another theme which runs through much of string theory is the search
for the symmetries which underlie the structure of string theory.  One symmetry
structure which has been investigated in this regard is that of
hyperbolic Kac-Moody
and generalized Kac-Moody algebras
\refs{\frenk, \golatt, \witthyper, \givpor, \moorei, \mooreii, 
\gebert,\gebnic} .
Generalized Kac-Moody (GKM) algebras occur very naturally
as unbroken symmetry groups in certain string
ground states \moorei\
and -contrary to what one expects in spontaneously
broken gauge theory - even when these
gauge symmetries are broken they nevertheless
put strong constraints on the
S-matrix \mooreii. The idea of a $T$-duality invariant
string algebra based on a Lorentzian lattice as a
gauge algebra for $N=4$ heterotic string compactifications
was  discussed  in
\givpor\ and is summarized in \givrev.
There are also close connections between these algebras
and the structure of string field theory \gnw.

In this paper we will provide evidence for a connection between
BPS states in  string theory with $N=2$ spacetime
supersymmetry and GKM algebras.
We will show that threshold corrections in $N=2$ heterotic string
compactifications are in fact determined purely in terms of the spectrum of BPS
states.  What is more surprising is that these corrections are closely
related to product formulae that have been studied recently by Borcherds,
Gritsenko and Nikulin \refs{\borchii, \borchiii, \nikulini, \nikulinii,
\nikuliniii }
in connection with generalized Kac-Moody
algebras.  We will show how this connection arises and
construct a GKM Lie superalgebra in terms of
vertex operators associated to BPS states.

Our results can be viewed
as a generalization to string theory of some of the structures
previously encountered in duality and $N=2$ Yang-Mills theory.
For example, our results suggest that the BPS states in string theory
should be regarded as ``gauge bosons '' of  the GKM algebra. We find
a direct generalization of the one-loop formula for the prepotential
in $N=2$ Yang-Mills theory \refs{\argf,\klti},
\eqn\one{\CF  = {i \over  4 \pi} \sum_{\vec \alpha>0} (\alpha\cdot A)^2
\log{(\alpha\cdot A)^2 \over  \Lambda^2 } }
where $A$ determines the components of the Higgs expectation
value in the Cartan sub-algebra $\phi = \sum A_i H_i $ and the
sum in \one\ runs over the positive roots of the Lie algebra of
the gauge group $G$.  In  a certain limit we will find
a similar formula in string theory where the sum over the positive roots of the
Lie
algebra of $G$ is replaced by a sum
over the positive roots of a GKM algebra.

The simplest example of a product formula which occurs in threshold
corrections is Borcherds' remarkable product formula for the
modular $j$ function
\eqn\jprod{ j(p) - j(q) = p^{-1} \prod_{n>0, m \in Z} (1-p^n q^m ) ^{c(mn)} }
where $j(q) -744 = \sum_{n=-1}c(n) q^n $.  The proof can be found
in \borcha and \borchi.  The infinite product converges
for $|p|$, $|q| $ $< e^{- 2 \pi}$ and is defined elsewhere
by analytic continuation.  An expansion of both sides for small $p$, $q$
quickly reveals that it requires an infinite number of identities among
the coefficients, the first of which is $c(4) = c(3) + c(1)(c(1) -1)/2$.
In \moonshine\ these identities appear as ``replication'' formulae involving
the dimensions of irreducible representations of the monster
group.  In \borchi\ the product formula \jprod\ is interpreted in
terms of the denominator formula for the monster Lie algebra
which is an example of a GKM algebra.  The quantity $j(T) - j(U) $
has appeared in the recent literature in connection with threshold
corrections to  gauge couplings with $T,U$ the moduli on a $T^2$
in the string compactification \refs{\clm, \dewit,\antoni }.
\foot{We use both notations $j(q)$ and $j(\tau)$, depending on
context.}
 In \refs{\clm, \dewit , \antoni }  this
term was found by indirect methods. We will show here how it and
various generalizations can be computed directly in terms
of products which generalize \jprod. Indeed, using the
integrals of appendix A it is straightforward to give an
independent proof of \jprod.

There is a substantial literature on threshold corrections
in $N=1$ and $N=2$ heterotic string theory \refs{\min,\vadim,\dkl,
\agn, \agnt, \antonii, \blst,\clm,
\antoni, \dewit, \mayrst, \kaplouis}.
 We will make particular use of some techniques of
 \dkl\ and of the result in \agn\kaplouis\
relating threshold corrections to the new supersymmetric index \nsi.
We also rely heavily on the results of
\refs{\dewit,\antoni}  relating explicit
one-loop string calculations to the quantities appearing in $N=2$
supergravity effective Lagrangians.
A connection between BPS states and threshold
corrections has been suggested, directly and indirectly,
in various  forms, in \refs{ \oogv, \fklz, \mayrst, \vaftest}.

The outline of this paper is as follows. In the second section we briefly
review the relevant properties of $N=2$ heterotic
string compactifications and associated moduli spaces.  In the
third section we discuss the
nature of perturbative BPS states in these compactifications.  We show
that threshold corrections are determined purely by the spectrum of BPS states
and relate these
corrections to the elliptic genus of the internal $ c=6 $, $N=4$ superconformal
field theory governing the compactification. In the fourth section we
determine the  dependence of these corrections on either the $(T,U)$
Narain moduli of $T^2$ or these plus the $E_8$ Wilson line moduli
and show that they are
naturally given by infinite sums of logarithmic functions.  The fifth
section contains a discussion of the relation of our work to a theorem
of Borcherds.  We discuss the automorphic structure of the products
which result from our analysis and show how Borcherds
results follow from an analysis of certain modular integrals.
In section six we discuss the quantum monodromy of the
prepotential.
The seventh section contains a brief
discussion of GKM algebras and their associated root
lattices and compares the results of sections four and five to the
denominator formula for these algebras. For the previous two choices of moduli
we show that the product formulae we obtain are given by a product
over positive roots of the monster Lie algebra and the hyperbolic
Kac-Moody
Lie algebra $E_{10}$ respectively. We discuss two limiting
cases of the prepotential in section eight, one of which
suggests that the GKM algebra should be considered a
gauge algebra in string theory.
In the ninth section we construct a generalized
Kac-Moody algebra which is in some sense associated to the BPS
states. The tenth section has some preliminary remarks
on the application of our results to $N=2$ string
duality. In the final section we conclude and offer some speculations
on possible extensions of this work.
Certain modular integrals needed in the evaluation of threshold
corrections are computed in an appendix.

\newsec{Review of $D=4,N=2$ heterotic string
compactifications}

\subsec{Chiral Algebra}

In this paper we will be considering compactifications of the
heterotic string to four spacetime dimensions with $N=2$
spacetime supersymmetry.   There is a $c=9$, $N=1$ internal superconformal
algebra (SCA) associated to any such compactification.  For such theories the
spacetime
supersymmetry implies that this internal SCA splits into a
 $c=6$ piece with  $N=4$ superconformal
symmetry  and a $c=3$ piece with $N=2$ superconformal symmetry \bksdx.
The $c=3$, $N=2$ theory  is constructed from two
free dimension $1/2$ superfields.   We will indicate this decomposition
of the right-moving superconformal algebra as
\eqn\chrlalg{
\tilde \CA^{N=2}_{\tilde c=3} \oplus \tilde
\CA^{N=4}_{\tilde c=6} \subset \tilde\CA \quad ,
}
where $\tilde \CA$ is the rightmoving chiral
algebra.
The left-moving internal conformal field theory has $c=22$
but is otherwise unconstrained except by modular invariance.

The $c=3$ theory has a $U(1)$ current which we denote by $J^{(1)}$. The
$c=6$ theory is in general non-trivial  and to be compatible
with $N=4$ superconformal symmetry must have a level one $SU(2)$ Kac-Moody
algebra.
Representations of the $c=6$ theory can thus be labelled by the
conformal weight and $SU(2)$ representation $(h,I)$.
We will also choose a $U(1)$ current
$J^{(2)}$ from this $SU(2)$ algebra with the
normalization $J^{(2)} = 2 J^3$ where $J^3$ is the $SU(2)$ Cartan current.
The total $U(1)$ current of the $c=9$ theory is $J= J^{(1)}+J^{(2)}$.

\subsec{Remarks on the moduli space}

$N=2$ heterotic compactifications
 typically have a large moduli space of vacua. The
moduli decompose into hypermultiplets and vectormultiplets under
the $N=2$ spacetime supersymmetry.  The $c=6$, $N=4$ SCA has
two massless representations in the Neveu-Schwarz
sector, $(h=0,I=0)$  and $(h=1/2,I=1/2)$ \egtaor. These
are associated to vectormultiplet and hypermultiplet moduli respectively.
The factorization \chrlalg\ is reflected in spacetime through the
constraints of $N=2$ supergravity which require a  local
factorization of the total moduli space of the form
\eqn\ntwosugra{
SK(n) \times Q(m)
}
where $SK(n)$ is the vectormultiplet moduli space which must
be a special Kahler manifold of
real dimension $2n$  and $Q(m)$ is the hypermultiplet moduli space and
is quaternionic of real dimension $4m$.
In fact the moduli space is a complicated and singular space;
its global structure has not been fully elucidated and
should prove very interesting.
The classical moduli space is an algebraic variety which probably
has the following structure. $\CM$  is a singular stratified space,
that is, there is a disjoint union
\eqn\strfd{
\CM = \amalg_{\cal s} \CM_{\cal s}
}
with smooth strata $\CM_{\cal s}$ which fit together in a singular
fashion. The $\CM_{\cal s}$ have
the structure of a fibration of a quaternionic
manifold over a special Kahler manifold.
\foot{In the case of global $N=2$ SYM with matter these
statements can probably be proven from the $F$ and $D$
flatness equations. For gauge group $G$ the strata
should be
 enumerated by the possible unbroken subgroups
$G_1 \subset G$. The base should be
$\liet(G_1)\otimes \IC /W(G_1)$
while the fibers are hyperkahler quotients.}

The classical vectormultiplet
moduli space for $N=2$ string compactifications is given by
the special Kahler manifold \fvanp
\eqn\fctr{
\CM_{\rm vm}^{s+2,2} \equiv
{ SU(1,1) \over U(1) }  \times \CN^{s+2,2}
}
where the  first factor is associated with the dilaton.
Here we have introduced the Narain moduli space
$ \CN^{s+2,2}$. This is a quotient of the generalized
upper half plane
\eqn\uhp{
\CH^{s+1,1} \equiv  O(s+2,2;\IR ) / [O(s+2)\times  O(2)]
}
by the left-action of the arithmetic subgroup
$ O(s+2,2;\IZ)$:
\eqn\narain{
\eqalign{
\CN^{s+2,2} & \equiv  O(s+2,2 ;\IZ)\backslash \CH^{s+1,1} \cr
& = O(s+2,2;\IZ)\backslash O(s+2,2;\IR ) / [ O(s+2) \times O(2) ]\cr}
}
The special geometry of \narain\ is described in some
detail below.

For $N=2$ compactifications with internal space $K3 \times T^2$
and an embedding of the spin connection in $E_8 \times E_8$ the
moduli space $\CM_{\rm vm}^{s+2,2}$ always contains a subvariety
$\cong \CN^{2,2}$,  which is the Narain moduli space for
the complex moduli $T,U$ on $T^2$:
\eqn\utpiece{ {O(2,2) \over O(2) \times O(2) } \simeq
\left( {SU(1,1) \over U(1) } \right)_T \otimes \left( {SU(1,1) \over U(1) }
\right)_U . }
The arithmetic group $O(2,2;\IZ)$ in this case consists of the $T
\leftrightarrow U$
exchange and transformations by $PSL(2,\IZ )_T \times PSL(2,\IZ ) _U $.

In $N=2$ string theory the classical special
geometry of $\CM_{\rm vm}^{s+2,2}$ receives quantum corrections.
The dependence of threshold corrections
on $T,U$ moduli has been well studied in the literature  and has
played an important role in recent tests of string-string
duality \refs{\klt,\klm,\agnti}.
In many compactifications based on $K3 \times T^2$ one finds additional
vectormultiplet moduli associated to Wilson lines for the unbroken
gauge group on $T^2$. The dependence of threshold corrections
on these Wilson line moduli has been studied in \refs{\clm, \dewit, \mayrsti}.

In this paper we will for the most part focus on two special cases with
moduli space \fctr\ for $s=0$ and $s=8$ corresponding to the $T,U$ moduli
on $T^2$ in the first case and in the second to $T,U$ and the Wilson
line moduli associated to the unbroken $E_8$ factor for $K3$ compactifications
given by the standard embedding of the spin connection in the gauge
group. For these two cases the moduli are just the Narain moduli
associated to the even self-dual lattices $\Pi^{2,2}$ and
$\Pi^{10,2}$.  However many of our considerations are more general and we
will only specialize to these two cases when necessary.

\subsec{The classical special geometry of $\CN^{s+2,2} $ and of
$\CM_{\rm vm}^{s+2,2}$. }

The standard Kahler geometry on the space \fctr\   is special geometry
described, for example, in
\refs{\dewit,\ferrarai}.
We now discuss a useful parametrization of the homogeneous
space
\eqn\cndef{ \CH^{s+1,1} = { O(s+2,2) \over O(s+2) \times O(2) } }
occurring in \fctr.
Further details may
be found in \refs{\cv, \gilmore, \dewit, \ferrarai, \clm}.

We can represent the homogeneous space
$\CH^{s+1,1}$ as a ``tube domain'' in a complexified
Minkowski space as follows.
Let  $\langle \cdot ,\cdot \rangle$ be a real quadratic form
of signature $(+^{s+2},-^2)$. Consider:
\eqn\tubei{
\{ u \in \IC^{s+4} : \langle u, u \rangle =0, \langle u, \bar u \rangle<0 \}
/(u\sim \lambda u)
}
This is a homogeneous space for
$O(s+2,2;\IR)$.
We let $y\in \IR^{s+1,1}\otimes \IC$, where we
have a real inner product $(,)$ on $\IR^{s+1,1}$ of
signature $(+^{s+1},-^1)$. We parametrize the solutions
to \tubei\ by
\eqn\tubeii{
 y \rightarrow u(y)\equiv \biggl( y;1, -\half (y,y) \biggr)
}
where  the inner products are related by
\eqn\tubeiii{
\langle v; x_1,x_2 \rangle^2 \equiv (v,v) + 2x_1 x_2
}
Moreover
\eqn\img{
\langle u(y) , \overline{u(y)} \rangle = +2 (\Im y, \Im y)
}
so we must have $(\Im y)^2<0$, i.e., a lightlike vector \foot{ We
will use both $\Re z$, $\Im z$ and $z_1$, $z_2$ to indicate the real and
imaginary parts of a complex number $z$ in this paper}.

The lightcone has two components, take
$\Im y \in C_+$,  the forward
light-cone, so we realize the moduli space as
a generalized upper half plane:
\eqn\tubeiv{
O(s+2,2)/[ O(s+2)\times O(2)] \cong
\CH^{s+1,1} \equiv  \IR^{s+1,1}+ i C_+^{s+1,1}
}

Sometimes when we want to be more specific about
the inner product in \tubeii\ we take
\eqn\specfip{
\eqalign{
v& = (\vec v; v_+, v_-) \cr
(v,v)& = \vec v^2 - 2 v_+ v_-\cr}
}
where $\vec v^2$ is the standard Euclidean
inner product.

One important special case is the submanifold
$\vec y=0$, isomorphic to the case $s=0$.
Then we have two complex numbers
$\Im y_+ \Im y_- >0$. The forward lightcone
then gives
\eqn\makeitup{
(y_+, y_-) \rightarrow (T,U) \in \CH\times \CH }
in the product of two upper half planes.

The Kahler metric on $\CN^{s+2,2}$ is
\eqn\khlr{
K  = - \log [ - \half \langle u, \bar u \rangle]
 = -\log [- (\Im  y^2 )  ]
 = - \log[ 2 \Im y_+ \Im y_-  - (\Im \vec y)^2  ] .
}
Note that the argument of the log is positive by the definition
of the domain of $y$ and that the resulting metric is one of
constant negative curvature. Similarly, the classical prepotential
and K\"ahler potential for the space \fctr\  are
$\CF = - S (y,y)$ and $K= - \log[8 \Im S] - \log[ -(\Im y)^2] $,
respectively.

For our purposes the space $\CN^{s+2,2}$ arises as the moduli space for Narain
compactifications based on the even self-dual Lorentzian lattice
$\Pi^{8t+2,2}$ when $s=8t$.
\foot{We use the symbol $\Pi^{p,q}$ to denote a
standard even unimodular Lorentzian lattice. Lattices
isomorphic to it are denoted by
$\Gamma^{p,q}(y)$, where $y\in \CN^{p,q}$.}
We write $\Pi^{8t+2,2} \cong \Pi^{8t,0} \oplus \Pi^{2,2}$
and write lattice vectors as
\eqn\fdintvii{
(\vec b; m_+, n_-; m_0, n_0)
}
with metric
\eqn\fdintviii{
(\vec b; m_+, n_-; m_0, n_0) ^2 = \vec b^2 -2 m_+ n_- + 2m_0 n_0 .
}
Here $\Pi^{8t,0}$ is an even self-dual Euclidean
lattice. For $t=0$ it is the zero lattice
$\{ \vec b=0\}$.
It is unique for $t=1$, there are two for
$t=2$, twenty-four for $t=3$, and so forth.

\newsec{Supersymmetric Index and BPS states
in $N=2$ heterotic string  compactifications}

\subsec{BPS States}

Compactifications of the heterotic string with $N=2$ spacetime supersymmetry
have a spacetime supersymmetry algebra which includes a complex central charge
$\CZ $:
\eqn\susytwo{  \{ Q_\alpha^i, Q_\beta^j \} = \epsilon_{\alpha \beta}
\epsilon^{ij} \CZ . }
The central charge $\CZ$ is determined by the right-moving momenta $p_R$
carried
by the free superfields in the $c=3$, $N=2$ internal SCA.
The Virasoro constraints
imply that the mass of any state in this theory is given by (in the
Neveu-Schwarz sector)
\eqn\two{M^2 = (N_R - 1/2) + \half p_R^2 + h_R }
where $N_R$ is the right-moving oscillator number coming from the
uncompactified coordinates and the two free superfields of
the $c=3$ theory and $h_R$ is the conformal weight in the
$c=6$ part of the internal SCA.  As a result,  bosonic BPS states which must
satisfy
 $M^2 = \half p_R^2 $
arise as right-moving ground states with either $N_R= 1/2$ and $h_R = 0$
or $N_R=0$ and $h_R= 1/2$.

For Narain compactifications with lattices $\Pi^{8t+2,2}$ and
lattice vectors \fdintvii\ we have
\eqn\fdintix{
\eqalign{
\half ( p_L^2 -  p_R^2) & = \half \vec b^2 - m_+ n_- + m_0 n_0\cr
\half p_R^2 & = {1 \over  -2 (\Im y)^2} \biggl\vert
\vec b \cdot \vec y -  m_+ y_-  -  n_- y_+ + n_0 -
 \half m_0 y^2 \biggr\vert^2 . \cr}
}
That is, the central charges are determined by the inner product
of the lattice vector \fdintvii\ and the coordinate $u(y)$.

In contrast to theories with $N=4$ spacetime supersymmetry, the spectrum
of BPS states in $N=2$ theories can have a ``chaotic'' nature,  that is
BPS states can appear and disappear under infinitesimal perturbations
of the moduli.  One simple example which illustrates this is a
symmetric  orbifold
limit of  $K3$. We consider  a $Z_2$ orbifold of a $\Pi^{4,4}$ lattice.
In the untwisted sector of the orbifold we will have states
labelled by momenta $(\tilde p_L, \tilde p_R ) \in \Pi^{4,4} $.
Such states can be BPS states only if $\tilde p_R = 0$. But the existence
of such states varies discontinuously as we vary the Narain moduli
associated to the $\Pi^{4,4}$ lattice.  For example if we focus
on one $S^1$ factor of radius $R$ then states with $(\tilde p_L \ne 0 , \tilde
p_R=0 )$
exist only for rational values of the modulus $2 R^2$.

This seems to be in contradiction with the usual wisdom that BPS states
must behave smoothly under perturbations of the theory. The resolution of
this puzzle is simply that these chaotic BPS states always appear in
hypermultiplet, vectormultiplet pairs. As one moves away from the special
points these BPS states pair into long representations of the $N=2$ spacetime
supersymmetry algebra and are no longer BPS saturated.   This also makes
it clear that threshold corrections cannot depend only  on  the number
 and charges of BPS states
if one is to obtain smooth functions of the moduli.  Rather, as we will
see in the following section, the threshold corrections depend only
on the difference between vectormultiplet and hypermultiplet BPS
states and this difference is a smooth function of the moduli.

\subsec{BPS states and the supersymmetric index}

It was shown in \agn\ that threshold corrections in
$N=2$ heterotic string compactifications  can be written
in terms of
the ``new supersymmetric index''  of \nsi:
\eqn\ccmn{
\Tr_{R} J_0 e^{i \pi J_0}
q^{L_0 - c/24}
\bar q^{\tilde L_0 - \tilde c/24} = {1\over  2 \pi i }
{\p \over  \p \theta} \vert_{\theta= \half }
\Tr_{\CH_{R}^{int}} q^{L_0 - c/24}
\bar q^{\tilde L_0 - \tilde c/24}e^{2 \pi i \theta J_0}
}
where the trace is over the Ramond sector
of the internal $(c,\tilde c) = (22,9)$  conformal field theory and $J_0$ is
the
total $U(1)$ charge defined earlier. In the literature $J_0$ is often
denoted by $F$.

We will now show that for $N=2$ compactifications one can
relate \ccmn\ to a sum over
BPS saturated states.  Using the decomposition \chrlalg
the Hilbert space of the internal
superconformal field theory  may be written as
\eqn\inlhs{
\CH^{\rm int} = \sum_{(p_R;h,I)} \CH^{(22,0)}_{p_R;h,I} \otimes
\tilde \CH^{(0,3)}_{p_R} \otimes \tilde \CH^{(0,6)}_{h,I}
}
where superscripts denote the Virasoro central
charges $(c,\tilde c)$,
the second factor is a Fock space for
the free N=2 superfield, and the third factor
is a unitary irrep of the $N=4$ SCA labelled by
the conformal weight and $SU(2)$ representation $(h,I)$.
Each summand in \inlhs\ is a tensor product of
representations of the $\tilde\CA_{N=2}^{\tilde c=3}$
and the $\tilde\CA_{N=4}^{\tilde c=6}$ algebras.
Accordingly, the $U(1)$ current may be decomposed
as  $J= J^{(1)} + J^{(2)}$ and hence we may rewrite
the trace on each summand in \inlhs\ as:
\eqn\ccmni{
\eqalign{
\Tr_{p_R\otimes (h,I) } & J_0 e^{i \pi J_0}
q^{L_0 - c/24}
\bar q^{\tilde L_0 - \tilde c/24} =
\qquad \qquad\qquad
\cr
 & \Tr_{p_R}
J_0^{(1)} e^{i \pi J_0^{(1)}}
\bar q^{\tilde L_0 - \tilde c/24}
\left( \Tr_{(h,I)}  e^{i \pi J_0^{(2)}}
q^{L_0 - c/24}
\bar q^{\tilde L_0 - \tilde c/24} \right)
\cr
 +
&  \Tr_{p_R}
e^{i \pi J_0^{(1)}}
\bar q^{\tilde L_0 - \tilde c/24}
\left( Tr_{(h,I)} J_0^{(2)} e^{i \pi J_0^{(2)}}
q^{L_0 - c/24}
\bar q^{\tilde L_0 - \tilde c/24} \right)
\cr}
}
Now, for an arbitrary $N=4$ representation
$(h,I)$  one has
\eqn\ccnfr{
\Tr_{(h,I)} J_0^{(2)} e^{i \pi J_0^{(2)}}
q^{L_0 - c/24}
\bar q^{\tilde L_0 - \tilde c/24}
=0 .
}
To see this recall that the $U(1)$ current is related to
the $SU(2)$ Cartan current by $J= 2 J^3$ and hence
has integral spectrum. Since in $SU(2)$ representations
eigenvalues of $J^3$ come in opposite pairs \ccnfr\ follows.
Thus, only the first term on the
right hand side of \ccmni\ survives.

We can further simplify \ccmni\  using N=4
representation theory.  As we saw earlier BPS states correspond
in the Neveu-Schwarz sector to operators in the $N=4$ theory
with $h_R = 0, 1/2$. These are just the massless NS representations
of the $N=4$ SCA with $(h=0,I=0)$ and $(h=1/2,I=1/2)$.  The
representation $(h=0,I=0)$ gives rise to the bosonic states in
a BPS vectormultiplet.  On the other hand the representation
$(h=1/2,I=1/2)$  gives rise to the bosonic
states of a BPS hypermultiplet\foot{Our use
of the terms ``hypermultiplet'' and ``vectormultiplet'' here is nonstandard.
We define these terms based purely on the right-moving
structure of the representation.  For example,
with this terminology the supergravity
 multiplet is counted as a ``vectormultiplet.''}.

Spectral flow in an  $N=4$ theory
maps these representations to  Ramond representations
according to
\eqn\nfrsf{
\eqalign{
 (h=0,I=0) \rightarrow (1/4,1/2) \cr
(h=1/2,I=1/2) \rightarrow (1/4,0) . \cr}
}
These are the massless Ramond representations
with Witten indices  \egtaor:
\eqn\wttindx{
\eqalign{
\Tr_{\CH^{N=4}_{(h=1/4,I=0)}} (-1)^{2 J^3_0} & =  1 \cr
\Tr_{\CH^{N=4}_{(h=1/4,I=1/2)} }(-1)^{2 J^3_0} & = - 2 \cr}
}
In fact, these are the only representations for which
the Witten indices are nonzero \egtaor. It follows that
the sum \ccmn\ reduces to a sum over BPS states.

The
representation content of BPS multiplets
with respect to   the subalgebra
$\tilde \CA^{c=3}_{N=2} \oplus \tilde \CA^{c=6}_{N=4}$
may be summarized as follows.
We will denote a representation of this algebra by $(h,q)\otimes (h',I)$
where $(h,q)$ give the conformal weight and $U(1)$ charge of the
$c=3, N=2$ theory and $(h',I)$ labels the representation
of the  $N=4$ algebra .
With this understood
vectormultiplets and hypermultiplets have the
following content in the Neveu-Schwarz (NS) and Ramond (R) sectors:
\eqn\vhcont{
\matrix{
  &  { \rm Vectormultiplets} & { \rm Hypermultiplets }  \cr
{\rm NS:} & 2 \times  (0,0)\otimes(0,0) \oplus (1/2,\pm 1)\otimes( 0,0) & 2
\times (0,0)\otimes(1/2,1/2 ) \cr
{\rm R:} &   (1/8,\pm 1/2)\otimes( 1/4, 1/2 ) &
2 \times (1/8,\pm 1/2)\otimes( 1/4, 0) \cr }
}
where all combinations of $\pm $ signs should be taken.

Combining \vhcont\
 with \wttindx\ we see that each BPS vectormultiplet
contributes
\eqn\vmcont{
[\half e^{ i \pi/2}  - \half e^{-i \pi/2} ] (-2) = -2 i
}
to $J_0 e^{i \pi J_0} $
 while each BPS  hypermultiplet contributes
\eqn\hmcont{
2 \times [\half e^{ i \pi/2}  - \half e^{-i \pi/2} ] (+1) = +2 i
}
to $J_0 e^{i \pi J_0}$.
Thus the BPS states contribute to \ccmn\
with vectormultiplets and hypermultiplets
weighted with opposite signs, that is:
\eqn\vmhm{
\eqalign{
{1 \over  \eta^2} \Tr_R J_0 e^{i \pi J_0}
q^{L_0 - c/24}
&
\bar q^{\tilde L_0 - \tilde c/24}
= \cr
- 2 i \Biggl[
\sum_{\rm BPS\  vectormultiplets}
q^{\Delta} \bar q^{\bar \Delta}
-&
\sum_{\rm BPS\  hypermultiplets}
q^{\Delta} \bar q^{\bar \Delta} \Biggr] \cr}
}
Strictly speaking this equation is not completely correct
because of the fact that non-physical states with $\Delta \ne \bar \Delta $
nonetheless contribute to modular integrals.  With the
understanding that the contribution of these  non-physical
BPS states should be included as well \vmhm\ is a correct
equation.

Note that this is in accord with the expectation
that there are no threshold corrections for
$N=4$ spacetime supersymmetry and indeed  $N=4$
BPS states split up into a $N=2 $ hypermultiplet
and a $N=2 $ vectormultiplet.
Another way to understand this result
from the spacetime point of view is to
notice that when representing the
extended supertranslation algebra the
massive {\it long} $N=2$ representations are
the same as the {\it short} $N=4 $ representations.
But we know there are never any threshold
corrections in $N=4 $ theories, therefore we expect
threshold corrections only from short $N=2$ representations,
that is from BPS states.

\subsec{Elliptic genus }

{}From \ccmni\ and \ccnfr\ we see that the new
supersymmetric index \ccmn\
depends on $N=4$ moduli only through
\eqn\ellnew{ \Tr_{\rm Ramond}  e^{i \pi J_0^{(2)}}
q^{L_0 - c/24}
\bar q^{\tilde L_0 - \tilde c/24}  }
which, roughly speaking, is the elliptic
 genus \refs{\schwarn, \witteg, \lercheeg } of the $N=4$ conformal field
theory. For a general class of backgrounds we can be more specific
about the relation to the conventional elliptic genera.
We assume that the gauge bundle on $T^2\times K3$
has the structure  $\pi_1^*(V_1) \oplus   \pi_2^*(V_2) $, that is,
 in the fermionic formulation of the gauge
algebra, the leftmoving fermions $\lambda^I$ can be
split into two disjoint sets coupling via their currents as
\eqn\fermcoup{
\lambda^I A^{IJ}_\mu \p X^\mu \lambda^J
+
\lambda^I B^{IJ}_\mu \p X^\mu \lambda^J
}
where $A$ is a flat connection on $T^2$ for the bundle $V_1$,
and $B$ is an anti-self-dual instanton  connection for the bundle
$V_2$ on $K3$.
We further assume that we can embed the connection $B$ in
an $SO(2n)$ subgroup of $SO(16) \subset E_8$.  We can then use
the bosonic formulation for one $E_8$ factor and the fermionic
formulation for the $E_8$ factor containing the  connection $B$
and split the $16$ fermions as $16 = 2n + (16 - 2n)$. We then couple the
$2n$ fermions to the connection $B$ via \fermcoup.
In this case we have $(16-2n) $ free Majorana-Weyl fermions coupled to
the flat connection $A$ on  $T^2$,
and a $(c, \tilde c)= (4+n,6)$
heterotic sigma model
on $K3$ with a gauge bundle
\eqn\ggebdle{
\matrix{V_2 \cr \downarrow \cr K3\cr}
}
satisfying the conditions
\eqn\ggebdlei{
c_1(V_2) = 0,  \qquad ch_2(V_2)=\half c_1(V_2)^2-c_2(V_2) = c_2(TK3) =+24 .
}
We now  define elliptic genera
for the $(4+n,6)$ sigma-model as
\eqn\elldefn{\eqalign{ \Phi^+(\hat A) & = Tr_{NS,R} (-1)^{F_R} q^{L_0 -c/24}
{\bar q}^{\tilde L_0 - \tilde c /24 } \cr
\Phi^- (\hat A) &= Tr_{NS,R} (-1)^{F_L + F_R} q^{L_0 -c/24}
{\bar q}^{\tilde L_0 - \tilde c /24 } \cr
\Phi(\Delta) &= Tr_{R,R} (-1)^{F_R} q^{L_0 -c/24}
{\bar q}^{\tilde L_0 - \tilde c /24 } \cr }}
where the subscript on the trace indicates the left-moving and right-moving
boundary conditions respectively and $F_L$ and
$F_R ( \equiv J_0^{(2)} ) $ are the left and right-moving
fermion number.   The $\Phi^\pm$ are the elliptic generalization
of the Dirac index \witteg\ where the $\pm$ determines whether odd
antisymmetric
tensor representations of $SO(2n)$ are counted with a plus sign or minus sign.
$\Phi^+$ is denoted by $H(q)$ in the first reference of \witteg.
The quantity $\Phi(\Delta)$  involves the elliptic generalization of the
Dirac index coupled to the spinor bundles associated to $V_2$.

We can combine these indices together with the remaining free
left-moving fermions,
summed over Ramond and Neveu-Schwarz boundary conditions to obtain
for the full trace
\eqn\exprell{\eqalign{
\Tr_{R} & q^{L_0 - c/24}
\bar   q^{\tilde L_0 - \tilde c/24} J_0 e^{ \pi i  J_0} \cr
 & = 2i  {Z_{10,2} \over \eta^{12}} \left[ ({ \vartheta_3 \over \eta} )^{8-n}
\Phi^+(\hat A) - ({ \vartheta_4 \over \eta} )^{8-n}
\Phi^-(\hat A) + ({ \vartheta_2 \over \eta} )^{8-n}
\Phi (\Delta)  \right]   \cr }}
where $Z_{10,2}/ \eta^{12}(\tau) $ is the partition function for
the free bosonic degrees of freedom on the $\Pi^{10,2}$ lattice constructed
from the $E_8$ lattice and the $\Pi^{2,2}$ lattice of $T^2$.

Since the elliptic genus depends only on the topology of the manifold
and the topology  of the
gauge bundle it is invariant under deformations of
the hypermultiplet moduli in accord with the decompositions
\ntwosugra\ and \chrlalg. Thus it can be evaluated by working
in a special limit of the hypermultiplet moduli such as an orbifold
limit.  For the standard embedding of the spin connection in
an $SU(2)$ subgroup of $E_8$ we can simply combine \exprell\ with
the results of \eguchi\ to obtain the answer.

\newsec{Threshold corrections and the Prepotential}

\subsec{General strategy}

In this section and the following we establish a connection between threshold
corrections in $N=2$ compactifications of
heterotic string theory and product formulae
studied recently by Borcherds in connection with  generalized
Kac-Moody algebras. This strongly suggests
the presence of  a GKM algebra in such compactifications. In the ninth section
we will discuss a construction of this GKM algebra in terms of vertex operators
for BPS states.

We will determine the prepotential
 by comparing
two formulae for one-loop renormalizations of nonabelian
gauge couplings\foot{In this section we follow the
``string-theoretic''
conventions of \kaplouis\dewit\ for moduli.
Thus, for example $\Re S>0$, $\Re T>0$,
etc. The conventions which are useful for
discussing automorphic properties are
related by
$$
y^{\rm string} = - i y^{\rm automorphic}
$$
}.

For a gauge group $G$ the one-loop coupling
renormalization is given by \kaplouis:
\eqn\thrshi{
\eqalign{
{1 \over  g^2(G;p^2) } & = k \Re\biggl[S + {1 \over  16 \pi^2} \Delta^{\rm
univ}\biggr] + {b(G) \over  16 \pi^2} \log{M_{\rm string}^2 \over  p^2}
+  {1\over  16 \pi^2} \Delta(G;y) \cr
 \Delta(G;y) =&
\int_{\CF} {d^2 \tau \over  \tau_2} \Biggl[ \CB- b(G) \Biggr] \cr}
}
where $\CB$ is given by a trace over the
internal Hilbert space:
\eqn\curlybee{
\CB=
- {i \over  \eta^2}
\Tr_{\CH^{int}_R} \biggl\{ J_0 e^{i \pi J_0}
q^{L_0 - 22/24}
\bar q^{\tilde L_0 -  9/24}\bigl[ Q^2 - {k\over 8\pi \tau_2}
\bigr]\biggr\}
}
Here $Q$ is a generator of the gauge group, $b(G)$ is the coefficient of the
one-loop beta function normalized as in \kaplouis\ and $S$ is the dilaton
field.
The ``universal'' term $\Delta^{\rm univ}$ is related to the ``Green-Schwarz''
term
which governs the one-loop mixing of the axion and moduli fields and plays
a role in the cancellation of sigma-model anomalies \refs{\dfkz, \lco,
\kaplouis}.
The quantity $k$ is the level of the Kac-Moody algebra associated to $G$.
Henceforth we set $k=1$.

{}From the discussion in the previous section it is clear that \curlybee\
receives
contributions only from BPS states. Morally speaking
\eqn\morals{ {1 \over g^2} \sim {1 \over 16 \pi^2} \bigl[
\sum_{vm} 2 Q^2 \log m^2 - \sum_{hm} 2 Q^2 \log m^2\bigr] }
in accord with the threshold rule of \dewit\ eqn. 3.45.

We can further elaborate \thrshi\ using
 the constraints of
 $N=2$ supergravity following the work of
\dewit.  In \thrshi\ we may express the
universal threshold correction in
terms of the one-loop correction to the
prepotential $h^{(1)}$:
\eqn\thrshiii{
{1 \over  16 \pi^2} \Delta^{\rm univ}
=
{1 \over - (\Re y)^2}
  \Re\Biggl[ \hone  - y_1^a {\p \over  \p y^a} \hone \Biggr]
}

On the other hand, using the Wilsonian
coupling  we may also write \kaplouis\dewit:
\eqn\thrshii{
\eqalign{
{1 \over  g^2(G;p^2) } & = \Re\biggl[\tilde S -
{1 \over  (s+4)  \pi^2}\log[\Psi(y)] \biggr]
  + {b(G) \over  16 \pi^2}
\biggl[ \log{M_{\rm Planck}^2 \over  p^2}
+ K(S,\bar S, y, \bar y ) \biggr] \cr}
}
where
\eqn\kahlr{
K=
- \log[\Re(S)] - \log[-(y_1)^2] + const.
}
and a subscript $1$ denotes the real part. The
Planck mass is given by $M_{\rm Planck}^2 = M_{\rm string}^2 \Re S $.
Since
we are using the string theory conventions
of  \refs{\kaplouis,\dewit}  for the moduli the real
part of $y$ is in the forward lightcone.
The coupling
$\tilde S$ differs from $S$ by the addition of
a holomorphic function such that $\tilde S$
is invariant under the duality group up to
shifts. In \dewit\ it is shown that we may
write
\foot{Note the change of sign from 4.30 in \dewit. We introduce
$\tilde S$ so that we can also discuss the
$(10,2)$ case. In the $(2,2)$ case it is related
to the invariant coupling $S^{\rm inv}$  of \dewit\ via:
$\tilde S = S^{\rm inv} -L/8 $. }:
\eqn\thrshiv{
\eqalign{
\tilde S  & = S +  {1 \over  s+4  }  \eta^{ab}{\p \over  \p y^a}
{\p \over  \p y^b} \hone \cr}
}
and under the perturbative duality group
we have
\eqn\almstinv{
\tilde S  \rightarrow \tilde S
+ {i \over 2(s+4) }\eta^{IJ} \Lambda_{IJ}
}
where $\Lambda_{IJ}$ is a real symmetric
matrix.

In \thrshii\ $\Psi(y)$ is a holomorphic  function
on $\CH^{s+1,1}$
such that
\eqn\autinv{
\Re\biggl[{1 \over  (s+4) \pi^2}
 \log \bigl[\Psi(y)\bigr] + { b(G) \over  16 \pi^2} \log\bigl[
-(y_1)^2\bigr]  \biggr]
}
is invariant under the perturbative
duality group $O(s+2,2;\IZ)$.
Requiring that the physical coupling
be free of any singularity on $\CN^{s+2,2}$
fixes the  divisor of $\Psi$.
These conditions determine $\Psi(y)$ up to an
overall constant, since if we have two such
then $\Psi_1/\Psi_2$ is a well-defined automorphic
function extending to the compactification
divisors of the variety $\CN^{s+2,2}$ and is
thus a constant.
\foot{Strictly speaking one must specify
appropriate asymptotic conditions to guarantee
uniqueness. However we do not need the
precise statement since we obtain the
prepotential by direct computation.}

Equating \thrshi\ and \thrshii\ gives a formula for
the prepotential. We now present the solution
for some special cases.
We will consider  a compactification of the heterotic
string on $K3 \times T^2$ with the standard embedding of the spin
connection in the gauge group.  This breaks the $E_8 \times E_8$ gauge group to
$E_7 \times E_8$. In the further reduction to four dimensions on
$T^2$ one can choose Wilson lines for the remaining unbroken
subgroup  leading at generic points to a low-energy gauge group
$U(1)^{s+4}$ with $s=15 $ with the $U(1)^4$ arising from the two
left-moving and two right-moving $U(1)$'s associated with the $T^2$
factor and the remaining  $U(1)^{15}$ corresponding to the Cartan
subalgebra of $E_7 \times E_8$.

When the unbroken gauge group is $[E_8 \times E_7 \times U(1)^2]_{left}\times
[U(1)^2]_{right}$ (or an enhancement
thereof)  we will compute the
running $E_8$ gauge coupling:
\eqn\egith{
{1 \over  g^2(E_8;p^2) } \qquad s=0
}
 as a function of the moduli $T,U$
parametrizing $\CN^{2,2}$.
When the gauge group is
$[  E_7 \times U(1)^{10}]_{left}\times [U(1)^2]_{right}$
we will compute
\eqn\sevne{
{1 \over  g^2(E_7;p^2) } \qquad s=8
}
as a function of the
moduli in $\CN^{10,2}$. The calculation for
both cases can be done in parallel.
The answer is given in section 4.4  below.

\subsec{Computing the integrand}

The first step in the calculation is
the evaluation of the integrand in
\thrshi.
As discussed in the previous
section,  the
 hypermultiplet dependence of the integrand
enters
only through the elliptic genus
and  therefore depends
only on the topology of the gauge bundle and the manifold
specifying the compactification. Thus, although
the integrand in principle requires evaluating
a partition function of a  super conformal field theory on K3,
we can perform the computation in an orbifold or other limit which is
smoothly connected to the theory of interest.
We will evaluate corrections by working in the
$T^4/Z_2$ orbifold limit of $K3$ with the standard orbifold
embedding of the twist in the gauge group.  Related computations
have been done  for the elliptic genus of $K3$ in \eguchi.

We start with the Narain lattice for a $T^6$ compactification to four
dimensions with the decomposition
\eqn\ndecomp{ \Gamma^{22,6} = \Gamma^{10,2} \oplus \Gamma^{4,4} \oplus
\Gamma^{8,0} }
We then mod out by an involution which acts as $-1$ on the $\Gamma^{4,4}$
factor
and as a shift $X^I \to X^I + \delta^I $ on $\Gamma^{8,0}$ with
\eqn\shft{
\delta=(\half, \half, 0^6) .
}

For this particular orbifold limit the unbroken gauge group at generic points
is in fact $U(1)^{20}$  and the vectormultiplet moduli space is
$\CN^{18,2}$.
However we will restrict ourselves in what follows to studying the dependence
of the threshold corrections on the moduli of the $\Gamma^{10,2}$ factor
in \ndecomp\  for which the classical moduli space is $\CN^{10,2}$. At
least for orbifolds it should be possible to study the dependence on other
moduli by generalizing the formulae in the appendix  to congruence subgroups
of $SL(2,Z)$.

We thus can evaluate:
\eqn\thrshv{
\eqalign{
\eta^{-2}(\tau)
\Tr_{R} q^{L_0 - c/24}
\bar   q^{\tilde L_0 - \tilde c/24} J_0 e^{ \pi i  J_0}
& = - 2i  Z_{10,2}
 {E_6 \over  \eta^{24}}  \hfil \cr
\eta^{-2}(\tau) \Tr_{R} q^{L_0 - c/24}
\bar   q^{\tilde L_0 - \tilde c/24} J_0 e^{ \pi i  J_0} [Q^2(E_7)
- {1 \over  8 \pi \tau_2} ]
&
=  - {i \over  12} Z_{10,2} {\cal E}_7  \cr
\eta^{-2}(\tau) \Tr_{R} q^{L_0 - c/24}
\bar   q^{\tilde L_0 - \tilde c/24} J_0 e^{ \pi i  J_0} [Q^2(E_8) - {1 \over  8
\pi \tau_2} ]
&  =  - {i \over  12} Z_{2,2}  {\cal E}_8 \cr  }}
where
\eqn\morthrsh{\eqalign{
Z_{s+2,2} & =  \sum_{p\in \Gamma^{s+2,2}}
q^{\half p_L^2} \bar q^{\half p_R^2} \cr
{\cal E}_7 & = \biggl(   {(E_2 - {3\over \pi \tau_2} )
 E_6 - E_4^2 \over  \eta^{24}}\biggl)  \cr
{\cal E}_8 & =  \biggl(   {(E_2 - {3\over \pi \tau_2} )
E_4 E_6 - E_6^2 \over  \eta^{24}}\biggl)   \cr    }
}
Here $E_{2n}(\tau)$ are the usual Eisenstein series and are given explicitly
in the appendix.  As a check note that  the first equation in \thrshv\ follows
from
\exprell\  and the results of \eguchi.

It follows that we can rewrite the coupling in \thrshi\ as
\eqn\thrshxx{
\eqalign{
{1 \over  g^2(G;p^2) } & = \Re\biggl[S + {1 \over  16 \pi^2}
\Delta^{\rm univ}\biggr] + {b(G) \over  16 \pi^2}
\log{M_{\rm string}^2 \over  p^2}   \cr
& - {1 \over  12} {1 \over  16 \pi^2} \biggl( \tilde \CI_{s+2,2} -
\CI_{s+ 2,2}
\biggr)   \cr}
}
Here we have introduced a class of integrals defined
in equations $(A.1), (A.2)$ of appendix A.
The specific integrals appearing in \thrshxx\
involve the modular forms:
\eqn\thrshvi{
\eqalign{
 s=8:\qquad {E_6 \over  \eta^{24}} & = \sum c_1(n) q^n =
q^{-1} - 480 + \cdots \cr
{E_4^2 \over  \eta^{24}} & = \sum c_3(n) q^n = q^{-1} + 504 + \cdots \cr}
}
\eqn\thrshvii{
\eqalign{
s=0: \qquad  {E_6 E_4 \over  \eta^{24}} & = \sum c_1(n) q^n =
q^{-1} - 240 + \cdots \cr
{E_6^2 \over  \eta^{24}} & = \sum c_3(n) q^n = q^{-1} - 984 + \cdots  \cr}
}
Note that the subtraction of the constant terms appearing
in the definition of the integrals $\tilde \CI , \CI$ and the
beta function are consistent for
\eqn\bfn{
\eqalign{
b(G) & = - {1 \over  12} ( \tilde c_1(0) - c_3(0) ) \cr
& = -60 \qquad s=0 \cr
& = +84 \qquad s=8 \cr}
}
In general  the coefficients $c_1(n)$ and $c_3(n)$ depend
both on the choice of gauge bundle and on which low-energy
gauge group is being studied.

\subsec{Formulae for the Integrals}

The integrals  $\CI , \tilde \CI$ are  evaluated in appendix A.
The integral $\CI$ is given by the expression:
\eqn\firstint{
\eqalign{
\CI_{s+2,2}(y)
& =- 2 \log \bigl\vert
\Phi(y) \bigr\vert^2
+ c_3(0) \biggl( - \log [-(\Re y)^2]   -{\cal K } \biggr)
 \cr
\Phi(y) & = e^{- 2 \pi \rho\cdot y}
\prod_{r>0 }
 \biggl(1-e^{- 2 \pi  r \cdot y  }
 \biggr)^{c_3(-r^2/2)}
\cr}
}
In the above ${\cal K}$ is a constant defined in the appendix.
The product over $r>0$ means the following.  We consider the even-self-dual
Lorentzian lattice $\Pi^{s+1,1}$ for $s=0,8$ and write lattice vectors as
$r = (\vec b, -\ell,-k )$ with $\vec b \in \Pi^{s,0}$.
For
$s=8$, $\Pi^{8,0}$ is the root lattice of $E_8$ and we
choose a set of positive roots for this lattice. With this
understood,
$r>0$ means:

1. $k>0$   or,

2. $k=0$, $\ell>0$, or,

3. $k=\ell=0$, $\vec b>0$.

For $s=8$ we show in section seven below that this is
the positive root condition for $E_{10}$. For $s=0$  the
coefficient $c_3(0)$ is not determined purely by modular invariance.
After explicitly
dividing out
 the terms involving $c_3(0)$ the remaining product
is over the positive roots of the Monster Lie algebra
\borcha\borchi.

The vector $\rho$ is:
\eqn\wvlvc{
\eqalign{
\rho =&  \rho_{E_{10}} = - (\vec \rho; 31, 30) \qquad s=8 \cr
\rho   = & - {c_3(0) \over  24} (1,1) + (0,1) \qquad s=0 . \cr }
}
where $\vec \rho$ is the Weyl vector of $E_8$.
For $s=8$, $\rho_{E_{10}}$
is the Weyl vector for $E_{10}$. For $s=0$  \wvlvc\ gives
a lattice Weyl vector for the $\Pi^{1,1}$ lattice but only gives the Weyl
vector for the Monster Lie algebra after subtraction of the $c_3(0)$ term.

The second integral we need is\foot{Integrals of this type are also
evaluated in \kirka\ although there are important differences
in our results.}:
\eqn\thrshxi{
\eqalign{
 \tilde \CI_{s+2,2}(y)  =
& 4\Re\Biggl\{
\sum_{r>0} \biggl[
 \tilde c_1(- {r^2 \over 2})
Li_1(e^{-2 \pi r\cdot y })
+ {6 \over  \pi ( y_1)^2} c_1(-{r^2 \over 2} ) \CP(i r\cdot y) \biggr]
\Biggr\}
 \cr
& + \tilde c_1(0) \Biggl( - \log [-( y_1)^2]
  - {\cal K}  \Biggr)
 + {1 \over  (y_1)^2} [ \tilde d^{s+2,2}_{ABC} y_1^A y_1^B y_1^C
+ \delta ]  \cr}
}
where
\eqn\condst{
\delta = {6 \over  \pi^2} c_1(0)  \zeta(3) .
}
Here $Li_1(x)= - \log(1-x)$ and  the function $\CP$ involves
the polylogarithms $Li_2$ and $Li_3$ and is defined in
the appendix.   As explained in section 5.1,
$\CH^{s+1,1}$ is tessellated by a system of
Weyl chambers, and the  symmetric tensor
$ \tilde d^{s+2,2}$ is piecewise constant in
$\CH^{s+1,1}$, taking different values in
different Weyl chambers. Explicit formulae
for it are given in appendix A.

\subsec{Answer for the prepotential}

Equating \thrshxx\ with \thrshii\  and using \thrshiii\ and \thrshiv\  we
obtain a
differential equation for $\hone$:
\eqn\diffleq{
\eqalign{
 \Re \biggl[ &
{1 \over  s+4  }  \eta^{ab}{\p \over  \p y^a}
{\p \over  \p y^b} \hone\biggr]
+
{1 \over  (\Re y)^2}
  \Re\Biggl[ \hone  - y_1^a {\p \over  \p y^a} \hone \Biggr] \cr
& =
- {1 \over  12} {1 \over  16 \pi^2}
\biggl( \tilde \CI_{s+2,2} - \CI_{s+ 2,2}
\biggr) + {1 \over  (s+4) \pi^2} \Re\bigl[ \log \Psi \bigr]  +
{b(G) \over 16 \pi^2} \log (-y_1^2)  .\cr}
}
A particular solution of the
differential  equation \diffleq, {\it in the
fundamental Weyl chamber},
\foot{See section 5.1}  is given by
\eqn\solution{
\hone  = {1 \over  384 \pi^2} \tilde d^{s+2,2}_{ABC} y^A y^B y^C
- {1\over 2 (2 \pi)^4} c_1(0) \zeta(3)
-  { 1 \over  (2 \pi)^4 }
  \sum_{r>0} c_1 (-r^2/2 )  Li_3( e^{ - 2 \pi  r\cdot y} )
}
Substitution into \diffleq\ leads to a solution with
\eqn\corrf{
\log \Psi(y) = \half \log[ J(i T) - J(i U) ] + b(E_8)
\log[\eta(iT) \eta(iU)] \qquad s=0
}
and
\eqn\corrsec{
\Psi(y) =
\Phi (y)  = e^{- 2 \pi \rho\cdot y}
\prod_{r>0 }
 \biggl(1-e^{- 2 \pi  r \cdot y  }
 \biggr)^{c_3(-r^2/2)} \qquad s=8 .
}

{\it Proof.}  We substitute directly the
ansatz  for $\hone$.  The universal term
is given by
\eqn\univtrm{
{1 \over  16 \pi^2} \Delta^{\rm univ}
= { 1 \over  y_1^2} { 1 \over  (2\pi)^3}
\Re\biggl[
\sum_{r>0} c_1(-r^2/2) \CP (i r \cdot y)
\biggr]+ {1 \over 192 \pi^2 (y_1)^2}
 [ \tilde d^{s+2,2}_{ABC} y_1^A y_1^B y_1^C
+ \delta ]
}
In order to cancel the other terms we must
use the identity:
\eqn\modident{
\eqalign{
\sum n c_1(n) q^n &= - {1 \over 2} {E_2 E_6 + E_4^2 \over  \eta^{24}}
\qquad \qquad \qquad s=8 \cr
\sum n c_1(n) q^n &= -{1 \over  6}  {E_2 E_4 E_6 + 2 E_6^2 +3 E_4^3
\over  \eta^{24}} \qquad s=0 \cr}
}

One can  check the rational terms directly, but
a more elegant method relates the trace of
$\tilde d_{ABC}$ to the Weyl vector. For
example, in the
case $s=8$ we proceed
as follows. The Laplacian for the
Kahler metric on $\CN^{s+2,2}$
is given by
\eqn\khlrlap{
\nabla^2  = -2 y_1^2\biggl( \eta^{ab} - {2 \over  y_1^2} y_1^a y_1^b
\biggr) \p_a \pb_{\bar b}
}
By straightforward computation one finds:
\eqn\lapeq{
\eqalign{
( \nabla^2 -24 ) \tilde \CI_{10,2} + 280 \tilde c_1(0)
& =
24 \biggl[ - 2 \log\vert \Phi(y)\vert^2 \qquad  \cr
& - c_3(0) (\log(-y_1^2) +{\cal K } ) + (8 \pi \rho_b
- {1 \over  4} \tilde d^a_{~ a b} )y_1^b \biggr]
\cr
}
}
since both sides of the equation must be
invariant under the duality group we have after
comparing to \firstint:
\eqn\trdee{
\tilde d^a_{~ a b} = - 32 \pi \rho_b
}
Similarly one finds
\eqn\ttlapl{
\eqalign{
( \nabla^2 -4 ) \tilde \CI_{2,2}
&
= 4 \tilde c(0)\bigl( \log\bigl[2 T_1 U_1
\vert \eta(i T)^4 \eta(i U)^4\vert\bigr] + \CK-\half ) \cr
& -80\log\vert J(iT) - J(iU)\vert\cr}
}
Using this equation one can check that
the rational terms in \solution\ solve
the equation required for equality of
\thrshi\ and \thrshii.

Above we have exhibited a particular
solution of the second order differential
equation. Two solutions
to this equation  must differ by a solution
of the homogeneous equation. It is
straightforward to show that the
only solutions of the homogeneous equation
which are analytic around zero are of the form:
\eqn\ambig{
\delta \hone = \sum_{0\leq m+n\leq 2} a_{IJ,mn}
(\hat X^I)^m (\hat X^J)^n
}
where the coefficients $a_{IJ,mn} $
are pure imaginary and
\foot{in a gauge where $X^0=1$}
\eqn\speccoord{
\hat X^I = (1, -y^2, i y^a).
}
Two prepotentials differing by such an
expression are physically equivalent.
We may use this result, together with the
automorphic properties discussed in
section six to argue that the prepotential
given above is the unique answer.

It is worthwhile to make several checks on
the above answer.
First, the physical coupling must be a
nonsingular function on moduli space.
We can check this since the
singularities of
$\log \Psi$ are  identical to
${1 \over  s+4} \bigl({\p \over  \p y}\bigr)^2 \hone$.
Second, the physical coupling must be
duality invariant. This is a consequence of
the automorphic properties
proved in the next two sections.
Third, in the case $s=0$, a differential
equation for the quantity
$S^{\rm inv} = S -
{1 \over  16 \pi^2} \Delta^{\rm univ}$
was derived in \refs{\kaplouis, \dewit}.
\foot{See, for example,
 equation  4.46 in \dewit.}
It is straightforward to show that
the above formula for $\hone$ satisfies
this constraint. As a fourth check we may
compare the ``Yukawa couplings''
\foot{Actually, magnetic moments,
\ferrarai.}
 following
from \solution\ with those derived in
\dewit\antoni. For example, one finds, in the $s=0$ case:
\eqn\yukawa{
\p_U^3 \hone = -{1 \over  2 \pi} \biggl[
1 - \sum_{r>0}c_1(k\ell) \ell^3
{ e^{ -2 \pi (k T+ \ell U) } \over  1-e^{ -2 \pi (k T+ \ell U) }}
\biggr]
}
According to \refs{\dewit,\antoni}\  \yukawa\
must coincide with
\eqn\yukawap{
-{1 \over  2 \pi} { E_4(iU) E_4(i T) E_6(i T) \over
(J(iT) - J(iU) ) \eta^{24}(iT) } \qquad .
}
Comparing the $T \rightarrow \infty$ and
$U \rightarrow \infty$ limits of the expressions
one finds perfect agreement. Expanding in power
series and comparing terms we find agreement to
$10^{th}$ order.

\subsec{Gravitational Corrections}

There are other terms in the effective supergravity
action which involve chiral densities and are therefore
computable to all orders of perturbation theory.
The first of these is the gravitational coupling
$F_1$ given by
\eqn\grvcrr{ F_1 =
{-i  \over  192 \pi^2}
\int_{\CF} {d^2 \tau \over  \tau_2}
\Biggl[ {1 \over  \eta^2}
\Tr_{R} J_0(-1)^{J_0}  q^{L_0 - 22/24}
\bar q^{\tilde L_0 -  9/24}  \biggl[ E_2 - {3\over  \pi \tau_2} \biggr]
- b_{grav} \Biggr]
}
where the $E_2$ factor arises from the
$Q_{grav}^2 = -2 \p_\tau \log \eta $ term in \agn.
 In heterotic string theory $F_1$ and its
generalizations $F_g$ can be computed at one-loop order and the
comparison between these calculations and genus $g$ computations
in dual twisted Calabi-Yau theories provide strong evidence for
$N=2$ string duality \refs{\agnti, \klt}.
Again it is clear
from the previous discussion that these quantities in heterotic
string theory receive contributions
only from BPS states.  There are also potential ambiguities in these
amplitudes which require a careful treatment of infrared divergences,
perhaps
along the lines presented in \kirka\ \foot{We thank J. Louis for pointing
this out to us.}.

\newsec{Comment on relation to work of Borcherds}

In this section we will discuss the automorphic properties
of the threshold corrections we have computed and compare
our results to work of Borcherds. We will work in ``automorphic''
conventions for the moduli as described in footnote 4 of
the previous section.

First we explain the relation between
Borcherds' ``rational quadratic divisors''
(RQD)
and enhanced symmetry points (ESP) in Narain compactifications.
Let $L\cong \Pi^{s+1,1}$ be an even unimodular
lattice, and consider the $\Pi^{s+2,2}$ lattice
$M\cong L \oplus \Pi^{1,1}$. We represent an
element of $M$ by  $v= (r;a_+,a_-)$ with $r \in L$ and  $a_\pm$
integer. In \borchii\ Borcherds defines
rational quadratic divisors to be the locus
$\CD(v)\subset \CH^{s+1,1}$ of
\eqn\rqd{
\langle (r;-a_+,a_-)  , (y; 1,-\half (y,y)) \rangle = 0
}
for
\eqn\posvt{
\langle (r;-a_+,a_-),(r;-a_+,a_-) \rangle
= r^2 - 2 a_+ a_- >0
}

{}From the  discussion in sec. 3.1 we know that BPS states are parametrized
by vectors $v=(r;a_+,a_-)\in \Pi^{s+2,2}\cong \Pi^{s+1,1}\oplus \Pi^{1,1}$
The central charge  of a BPS state with quantum numbers
$(r;a_+,a_-)\in \Pi^{s+1,1}\oplus \Pi^{1,1} $ is just the
inner product \clm
\eqn\cntrl{
{\CZ}(v,y) = (r,y) + a_- -  \half a_+ y^2
= \langle(r;a_+,a_-), (y;1,-\half y^2) \rangle
}
Moreover, for such a state
\eqn\lftright{
p_L^2 - p_R^2
= r^2 + 2 a_+ a_-
}
Thus in  string theory only the
divisors with $\langle (r;-a_+,a_-),(r;-a_+,a_-) \rangle=2$
are of importance and these correspond to enhanced symmetry
points in the Narain moduli space.

In \borchii\borchiii\ Borcherds has proved the following
theorem:

{\bf Theorem}. Let $f(\tau)=\sum c(n) q^n$ be
a meromorphic modular form with all poles at
cusps. Suppose that $f$ is of weight
$-s/2$ for $SL(2,\IZ)$ and has integer coefficients, with
$24 | c(0)$ if $s=0$.
Then there is a unique vector $\rho\in L$ such that
\eqn\borprod{
\Phi(y)=e^{2 \pi i \rho \cdot y } \prod_{r>0,r\in \Pi^{s+1,1} } (1-e^{2 \pi i
r\cdot y })^{c(-r.r/2)}
}
can be analytically continued to define
a meromorphic automorphic form of weight
$c(0)/2$ for $O(s+2,2;\IZ)^+$. All the zeroes and poles
of $\Phi$ lie on rational quadratic divisors, and the
multiplicity of the zero of $\Phi$ at the rational quadratic divisor of the
triple
$(b;a_1,a_2)$ is
\eqn\borsome{\sum_{n>0} c(n^2(a_1 a_2 - (b,b)/2)) }

In \borprod\ $r>0 $ means that $r$ has positive inner product
with a chosen negative norm vector in $L$.
In some cases the product
\borprod\ is known to be the denominator formula for a generalized Kac-Moody
algebra.

We have seen  that products of precisely this
type  arise in the analysis of threshold
corrections in $N=2$ heterotic string compactifications.
Indeed, by studying the properties of the integrals
$\CI_{s+2,2}(y)$ of appendix A
we can easily rederive many of the results
discovered by Borcherds.

To do this
we must explain
the holomorphic factorization of the product arising
in the expression for $\CI$ in equation $(A.29)$ of appendix A.
In the previous section we
presented the results only in a particular Weyl chamber
(defined below). However the integrals as evaluated in the
appendix depend
on the ``hatted'' dot product $r \hat \cdot y$
rather than  the dot product $r \cdot y$.
The quantity $r \hat \cdot y$ is defined by:
\eqn\nnotation{
\hat{r \cdot y}  \equiv \Re \biggl[
(\vec b \cdot \vec y
+ \ell y_- + k y_+)\biggr]
+ i \biggl\vert  \Im \biggl[
(\vec b \cdot \vec y
+ \ell y_- + k y_+)\biggr] \biggr\vert
}
when $k>0$, and, when $k=0$:
\eqn\nnotationp{
\hat{r \cdot y}  \equiv
r\cdot y - N(r,y) y_-
}
where
\eqn\dummya{ N(r,y) = \sgn(\vec b) \biggl[ \sgn(\vec b)
{\vec b \cdot \Im \vec y \over
y_{-,2} } \biggr]
}
is an integer ($[ \cdot]$ is the greatest integer function.)
In order to understand  the holomorphic factorization
we must introduce Weyl chambers.

\subsec{Weyl chambers}

We can tessellate the forward lightcone
$C_+^{s+1,1}\subset \CH^{s+1,1}$
into
convex polyhedra whose walls are
defined by the
{\it real} codimension one
subvarieties
\eqn\margst{
\Im (r\cdot y ) = 0
}
for $r^2 =2, r\in \Pi^{s+1,1}$. These
are the ``surfaces of marginal stability''
and must be crossed when circling
the divisors of vanishing BPS central
charges.

Given a choice of simple roots $r_\mu$ for
the lattice $\Pi^{s+1,1}$ we can define
a  {\it fundamental Weyl
chamber} by the equation
\eqn\fndwc{
{\CC}{\CW}_0 \equiv \{ y: r_\mu \cdot \Im y >0 \}
}

Explicit choices of simple roots are described in section
seven.
For $s=0$ there are two Weyl chambers
and we take the
fundamental Weyl chamber to be
simply  $\ypt > \ymt $.
For $s=8$ we choose a set of positive roots
 $\vec b>0$, $\vec b^2 =2$ for $E_8$.
Using the simple roots in the next
section we get the conditions:
\eqn\intrvl{
\eqalign{
0 <  & {\vec b \cdot \Im \vec y \over  \ymt}< 1\cr
  \ymt &  < \ypt \cr}
}

\subsec{Holomorphic factorization and $\CI_{s+2,2}(y)$}

By definition of the fundamental Weyl
chamber we know that $\hat{r\cdot y}= r\cdot y$
for the positive roots. Thus, the products
$(A.31),(A.37)$
factorize straightforwardly in the fundamental
chamber. The question arises as to the
other chambers.

Contributions from $c(n)$ for $n>0$
always holomorphically factorize.
To see this note that
$\Im y$ is  in the forward light cone so
\eqn\frwrd{
\eqalign{
\Im y_+ & > 0 \cr
\Im y_- & > 0 \cr
2 \Im y_+ \Im y_-  & > (\Im \vec y)^2 \cr}
}
Since we are studying coefficients
$c(n)$ for $n>  0$ in the product
we have $2 k \ell >( \vec b)^2 $.
It follows that
\eqn\dummyb{
(k \Im y_+ + \ell \Im y_-)^2 > 4 (k \Im y_+) (  \ell \Im y_-)
> \vert \vec b \cdot \Im \vec y \vert^2
}
so the imaginary part of $r\cdot y$ is always
positive and the product is a holomorphic square.
Similarly, the coefficient
$c(0)$ enters for roots
$k=0, \ell>0, \vec b=0$ and for roots
$k\ell=\half \vec b^2>0$. For these roots
we can again drop the hat. All of this is
quite simply understood from the physical
point of view: only states with $v^2=2$ can
become massless and lead to nonanalytic
behavior.

The real source of nontrivial holomorphic
factorization is entirely in the coefficients
of $c(-1)$ which are connected to RQD's
or ESP's. For example, the peculiar shift
in \dummya\ affects these terms. The
key observation is that these changes
can be absorbed in a change of the
linear term  $\rho\cdot y \rightarrow \rho_\alpha\cdot y$
in the RHS of $(A.29)$
in each Weyl chamber $\CC_\alpha$.
Consider for example,
 the holomorphic factorization
for roots with $r^2=+2$ and $k>0$. Then:
\eqn\holfcti{
\log\bigl\vert 1-e^{2 \pi i \hat{r \cdot y}  }
 \bigr\vert^2 =
\cases{
\log\bigl\vert 1-e^{2 \pi i  r \cdot y }
 \bigr\vert^2  & for $\Im r \cdot y>0 $ \cr
\log\bigl\vert e^{- 2 \pi i  r \cdot y }(1-e^{2 \pi i  r \cdot y })
 \bigr\vert^2  & for $\Im r \cdot y<0 $ \cr
}
}
For example, consider the case $s=0$.
There is only one positive root with
$r^2=2$, namely $r=(\ell,k)=(-1,1)$. The
wall is just
\eqn\walli{
\Im y_+ = \Im y_-
}
i.e. $T_2=U_2$.
Consequently the product in \firstint\ becomes
\eqn\wallii{
-2 \log\biggl\vert e^{-2 \pi i T c(-1) } \prod_{r>0}
\Biggl(1-e^{2 \pi i (k T+ \ell U) }
\biggr)^{c(k\ell)}
\Biggr\vert^2
}
for $T_2>U_2$ and
\eqn\wallii{
-2 \log\biggl\vert e^{-2 \pi i U c(-1) } \prod_{r>0}
\Biggl(1-e^{2 \pi i (k U+ \ell T) }
\biggr)^{c(k\ell)}
\Biggr\vert^2
}
for $U_2>T_2$.

\subsec{Automorphic properties of $\Phi$}

We will now use
invariance of the integral under
the group generated by

1. $y \rightarrow y + \lambda$ , $\lambda\in \Pi^{s+1,1}$

2. $y\rightarrow w(y)$, $w\in O(s+1,1;\IZ)^+$

3. $y \rightarrow y' = 2y/(y,y)  $

\noindent to deduce some automorphic properties of
the ``holomorphic square root'' $\Phi(y)$.
It is evident that the first two transformations
act on $\CH^{s+1,1}$. The third is surprising,
but note that under this transformation:
\eqn\chgimy{
(\Im y' , \Im y') = +{4 \over  \vert (y,y)\vert^2} (\Im y, \Im y)
}
and so preserves the ``upper half plane.''
In fact, the transformations 1.,2. and 3. only generate an
index 2 subgroup of the duality group, and
$\Phi$ is automorphic for the entire group
$O(s+2,2;\IZ)$.

Invariance under $y \rightarrow y + \lambda$
is trivial.
To verify $y\rightarrow w(y)$ invariance we
use \holfcti\ which says that in a Weyl
chamber $\CC_\alpha$ we have
\eqn\othchmb{
\CI = -2 \log \vert e^{(\rho_\alpha - \rho) \cdot y} \Phi(y)
\vert^2 + c(0) \biggl( - \log [-(\Im y)^2]   - {\cal K } \biggr)}
Now we note that
\eqn\othchbi{
(\rho_\alpha - \rho) = w(\rho) - \rho
}
To prove this note that it is true for chambers
related to the fundamental chamber by
a reflection in a simple root $r_i$. Then
proceed from there by induction.

The transformation of $\Phi$ under the
third generator is the most difficult to prove
by straightforward methods. We may
note that if $\Re y =0$ then the transformation
is a rescaling of $\Im y$ by a positive
coefficient. Therefore, an open set of the
fundamental Weyl chamber is mapped to
itself.   From the invariance of the
integral $\CI_{s+2,2}$ and its holomorphic factorization
we then find the modular transformation law:
\eqn\modtmn{
\Phi(y') = \bigl[ {(y,y) \over  2} \bigr]^{c(0)/2} \Phi(y)
}

\subsec{Zeroes and Poles of $\Phi$}

Finally, we come to the last part of Borcherds'
theorem which is concerned with the
zeros and poles of $\Phi$.
Unlike the walls of the Weyl
chambers, which are determined by
$r\in \Pi^{s+1,1}$, the RQD's are determined
by vectors $v\in \Pi^{s+2,2}$ with $v^2=2$.
As $y\to \CD(v)$ for some vectors
in the Narain lattice $\Gamma^{s+2,2}(y)$,
$p_R\rightarrow 0$, $p_L^2 \rightarrow 2$.
Each such vector contributes
 a logarithmic divergence to $\CI_{r+2,2}(y)$:
\eqn\divi{
\int^\infty {d\tau_2 \over  \tau_2} e^{-4 \pi \tau_2 \half p_R^2}
\rightarrow -  \log \vert \CZ(v,y)\vert^2
}

{\it Analysis of RQD's: (2,2) case}

The RQD's in $\CH^{1,1}$ are given by
the divisor $T=U$ and modular transformations
of it. These have a double intersection at
the points $T=U=i$
and a triple intersection at $T=U=\rho$
(and their modular images).
This is in accord with standard analysis of the
enhanced symmetry points.

One proof of the above statements proceeds
by writing:
\eqn\rqdi{
\CZ = \pmatrix{y_+ & 1 \cr}
\pmatrix{n_0 & n_- \cr m_+ & m_0\cr} \pmatrix{ y_- \cr 1\cr}
}
where $r^2=2$ implies $m_0 n_0 - m_+ n_-=1$.
Note that $SL(2,Z)\times SL(2,Z)$ acts on
the sets of divisors. By acting on $y_-$ we can
transform the matrix to  the unit matrix, so the
divisor equation becomes $1+ y_+ y_- =0$ or,
by a further
$SL(2,\IZ)$ transformation $y_+=y_-$.
We obtain the set of all divisors in $\CH\times\CH$
by taking $SL(2,Z)\times SL(2,Z)$ images of
this divisor. Note that divisors can only intersect
when both $y_+$ and $y_-$ are fixed by
some element of $SL(2,\IZ)$. In particular,
the divisor $y_+=y_-$  has a double intersection
at $i$ and a triple intersection at $\rho$.

{\it Analysis of RQD's: (10,2) case}

It would be interesting to give a classification
of the RQD's in $\CH^{9,1}$ analogous to
that above.
An important class of RQD's are
defined by the
simple roots of $E_8$.
For example, define the hyperplane
\eqn\splroot{
\CH_i = \{ y: \vec y \cdot \vec \alpha_i =0 \}
}
Along this hyperplane $\Phi$ has a simple zero
\eqn\splzer{
\Phi (y) \rightarrow - 2 \pi i \vec \alpha_i \cdot \vec y
\Phi' (y')
}
where, for $y'\in \CH_i$ we have
\eqn\newprod{
\Phi' (y')
 =
e^{2 \pi i \rho' \cdot y'} \prod'_{r'>0} (1- e^{2 \pi i r' \cdot y'}
)^{c'_3(-r^2/2)}
}
where $r'\sim r $ if $r-r' \propto \alpha_i$ and
$c'_3(-r^2/2)= \sum_{r' \sim r} c_3(-(r' )^2/2)$.
Restricting to successive intersections of divisors
we get a series of automorphic forms beginning with
the $E_{10}$ form $\Phi$ of \firstint\
and ending with  $J(T) - J(U)$.

\newsec{Quantum Monodromy}

\subsec{General Remarks}

The nature of the semiclassical monodromy of
$N=2$ heterotic string compactifications has been
thoroughly analyzed in references \refs{\dewit,\antoni}.
Using the above explicit expressions we can
compute the monodromy directly.

As discussed in \refs{\ferrarai,\dewit,\antoni} the best
basis of special coordinates for
questions of monodromy is the basis
$\hat X^I$ related to the basis $S,y^a$
with prepotential $\CF = (X^0)^2 S (y,y)$
by a duality transformation $\hat X^1 = F_1,
\hat F_1 = - X^1$.
Explicitly we have:
\eqn\hattd{
\eqalign{
\hat X^1 & = X^0 (y, y)^2\cr
\hat X^a & = X^0 y^a \qquad a=2,\dots s+3\cr}
}
In this basis
 the general transformation rule
for the one-loop prepotential, in automorphic
conventions is:
\eqn\gentmn{
\hone(\tilde y) = \biggl({\tilde X^0 \over  X^0}\biggr)^{-2} \biggl[
\hone(y) - i \Lambda_{IJ} {\hat X^I\over  X^0}
{\hat X^J\over  X^0}\biggr]
}
where $\Lambda_{IJ}$ is a real symmetric matrix.
Thus, the prepotential is an automorphic form of
weight $-2$ transforming with shifts.

In discussing monodromy one should always bear in
mind that the prepotential is ambiguous by an
addition of a term of the form \ambig.

\subsec{Monodromies for $s=0$}

It is convenient to introduce the function
\eqn\ellfun{
\eqalign{
\CL(T,U) & \equiv
 \sum_{r>0 } c(-r^2/2) Li_3(e^{2 \pi i (k T + \ell U )}) \cr
& =
Li_3(e^{2 \pi i  (T-U)}) +
 \sum_{k,\ell \geq 0} c(k\ell) Li_3(e^{2 \pi i (k T + \ell U )}) \cr}
}
where
\eqn\solprpi{
F(q) = \sum_{n=-1}^\infty c(n) q^n =  {E_4 E_6 \over  \eta^{24}} \quad .
}
This function has a branch locus at $T=U$ and
can be defined by power series for $T_2>U_2$, i.e.,
in the fundamental Weyl chamber.

The function $Li_3$ can be analytically continued
outside the unit circle and satisfies the connection
formula  \levin:
\eqn\monodr{
Li_3(e^x ) =
Li_3(e^{ - x }) + {\pi^2 \over  3} x
- {i \pi \over  2} x^2 -{1 \over  6} x^3
}
Thus, under analytic
continuation into the other Weyl chamber with
$U_2>T_2$ we have:
\eqn\ellrgii{
\CL(T,U) = \CL(U,T) - (2 \pi)^4{i \over  24 \pi} \bigl[
2U^3 - 2 T^3 - 3(T-U)^2 + (T-U)(6 T U -1) \bigr]
}
Under the generators $T\rightarrow T+1$ and
$U \rightarrow U+1$ of the duality group the function
$\CL(T,U)$ is invariant in its region of convergence.
Finally, to compute the monodromy under
$T \rightarrow -1/T$ (and similarly for $U \rightarrow -1/U$)
we rewrite the integral
$\tilde \CI_{2,2}$
as:
\eqn\rwrtl{
\eqalign{
\tilde \CI_{2,2}  = - { 6 \over U_2 T_2 \pi^2}
 &
\Re \Biggl[
\biggl( 1- i U_2 {\p \over  \p U} \biggr)
\biggl( 1 - i T_2 {\p \over  \p T} \biggr)\CL(T,U)
\Biggr]\cr
+ 20 \log\vert J(T) - J(U) \vert &
+ 264 [\log[2 T_2 U_2 \vert \eta(T) \eta(U)\vert^4 ]-\CK] - {\delta \over  2
T_2 U_2} \cr
&  + 16 \pi {U_2^2 \over  T_2} \qquad T_2\geq U_2 \cr
&  + 16 \pi {T_2^2 \over  U_2} \qquad U_2\geq T_2 \cr}
}
We know the integral $\tilde \CI_{2,2}$ is invariant
under $T \rightarrow T' = -1/T, U \rightarrow U$ and from this is
it straightforward to derive the transformation law:
\eqn\ellinv{
\eqalign{
\CL(T' , U) & = T^{-2} \biggl[ \CL(T,U) + \CR(T,U) \biggr] \cr
\CR(T,U) & = { \pi^2 \over  12} \delta (1-T^2) -
{4 \pi^3 i \over  3}  U^3 (T^2-1) \qquad \qquad\quad  T_2' > U_2 \cr
& = { \pi^2 \over  12} \delta (1-T^2) -
{4 \pi^3 i \over  3}  ({ 1 - T^4 \over  T})+
{4 \pi^3 i \over  3} U^3
 \qquad T_2' < U_2 \cr}
}

In the 2,2 case there are two Weyl chambers,
and accordingly we have the two expressions
for the prepotential (in automorphic conventions):
\eqn\solprep{
\eqalign{
\hone & = -{1 \over  (2 \pi)^4 }\CL(T,U) - { \delta \over  192 \pi^2}
- i {U^3 \over  12 \pi} + \delta \hone_+ \qquad T_2 > U_2 \cr
& =  -{1 \over  (2 \pi)^4 }\CL(U,T) - { \delta \over  192 \pi^2}
- i {T^3 \over  12 \pi} + \delta \hone_- \qquad T_2 < U_2 \cr}
}
where
$\delta \hone_\pm$ correspond to the
ambiguity \ambig.

Using these expressions and the above transformation
rules for $\CL(T,U)$ it is straightforward to compute
the monodromy of $\hone.$

\subsec{Remarks on the $s=8$ case}

In principle the above discussion generalizes
in a fairly straightforward way to the computation
of the monodromy on $\CN^{10,2}$. The Weyl
reflections $\sigma_i$ in the $E_{10}$ roots
no longer square to zero because of the
monodromy associated with the $Li_3$ terms
around enhanced symmetry varieties.
The transformation of $\CL(y)$ under the
inversion $y \rightarrow 2 y/ (y,y) $
can be deduced from an
analog of \rwrtl\ for $\tilde \CI_{10,2}$.
We hope to return to a detailed discussion of
this monodromy representation, which
gives an interesting representation of the
$E_{10}$ braid group, in the future.

\newsec{GKM algebras and denominator formulae}

In this section we will provide a very brief summary of  some of
the properties of GKM algebras and the hyperbolic
Kac-Moody algebra $E_{10}$ which enter into the interpretation
of the product  formulae for threshold corrections found in the
previous section. For further details on Kac-Moody algebras
and GKM algebras the reader can consult
\refs{\borcha, \borchalg, \kac, \fuchs, \gebert }.  $E_{10}$ is
discussed in \refs{\bjulia, \kmw, \gebnic, \gebnici, \gnw}

We first give the formal definition in terms of Cartan matrices,
generators and relations. Recall the usual definition of
a finite-dimensional simple Lie algebra. One starts with a symmetric
$r \times r$ Cartan matrix $A= (a_{ij}) $ with $i,j=1,2, \cdots r $
satisfying
\eqn\liedef{\eqalign{ a_{ii} & = 2,  \cr
     a_{ij}  & \le 0 \quad {\rm for } \quad  i \ne j , \cr
     a_{ij} & \in Z , \cr } }
and
$ {\rm det} A > 0$.
The Lie algebra is then defined in terms of generators $(e_i, f_i, h_i)$
obeying the relations
\eqn\lierel{\eqalign{ [ h_i, h_j ] & = 0 , \cr
                     [h_i, e_j ] & = a_{ji} e_j, \cr
                     [h_i, f_j ] & = - a_{ji} f_j, \cr
                     [e_i, f_j ] & = \delta_{ij}  h_j \cr
                   ({\rm ad}_{e_i})^{1-a_{ji}} e_j & = 0, \cr
                   ({\rm ad}_{f_i})^{1-a_{ji}} f_j & = 0. \cr } }
A formal construction of Kac-Moody Lie algebras is obtained by
weakening the condition that $\det A>0$. In particular
one obtains affine Lie algebras by allowing $A$ to have a zero
eigenvalue and hyperbolic Kac-Moody algebras by allowing
 a single negative eigenvalue.  For Kac-Moody algebras one
has the usual notion of root spaces, positive roots, simple roots,
and the Weyl group generated by reflections in the real simple roots.
Furthermore there is a denominator formula
\eqn\denform{e^\rho \prod_{r>0} (1-e^r)^{{\rm mult}(r)} =
    \sum_{\sigma \in W}  (\sgn (\sigma) ) e^{\sigma (\rho) } }
where $\rho$ is the Weyl vector, the product on the left hand
side runs over all positive roots and each term is weighted
by the root multiplicity ${\rm mult}(r)$, and on the right hand side
the sum is over all elements of the Weyl group $W$.  For example
for the $su(2)$ level one affine Lie algebra \denform\ is just the
Jacobi triple product identity.

Generalized Kac-Moody algebras have a great deal in common
with Kac-Moody algebras of hyperbolic type but differ from them
in that simple roots $r$ with $r^2 \le 0 $ are allowed.  The formal definition
follows from a slight generalization of the conditions \liedef.
One again starts with a symmetric Cartan matrix   $A= (a_{ij}) $
which is allowed to have infinite rank in general.  Then one demands
the conditions
\eqn\gkmdef{\eqalign{ a_{ii} & = 2 \quad  {\rm or}
        \quad a_{ii} \le 0,  \cr
         a_{ij}  & \le 0 \quad {\rm for } \quad  i \ne j , \cr
         a_{ij} & \in Z \quad {\rm if} \quad a_{ii} =2, \cr } }
and defines the algebra by generators $(h_{ij}, e_i, f_i)$ and
relations
\eqn\lierel{\eqalign{[h_{ij}, e_k ] & =
       \delta_{ij}a_{jk} e_k , \cr
       [h_{ij}, f_k ] & = - \delta_{ij} a_{jk} f_k, \cr
         [e_i,f_j ] & = h_{ij} \cr
        ({\rm ad}_{e_i})^{1-a_{ji}} e^j &= 0, \quad
      ({\rm ad}_{f_i})^{1-a_{ji}} f^j = 0,
            \quad a_{ii}=2, i \ne j , \cr
      [e_i,e_j] & =0, \quad [f_i,f_j] =0,
            \quad  a_{ii} \le 0, a_{jj}<0, a_{ij}=0 . \cr} }
The fact that $a_{ii}\le 0 $ is allowed shows that imaginary simple roots
($r^2 \le 0$) appear in contrast to Kac-Moody algebras.
There is again a denominator formula for GKM algebras which
reads
\eqn\dengkm{e^\rho \prod_{r>0} (1-e^r)^{{\rm mult}(r)} =
    \sum_{\sigma \in W}  (\sgn (\sigma) ) \sigma ( e^\rho \sum_r \epsilon(r)
e^r ) }
where the correction factor on the right involves $\epsilon(r)$ which is
$(-1)^n$ if $r$ is the sum of $n$ distinct pairwise orthogonal imaginary roots
and zero otherwise.

As with hyperbolic Kac-Moody algebras, these formal definitions
are of little practical utility without some method of determining the root
multiplicities.
For a few special GKM algebras
it is  possible in some cases to determine root multiplicities
through the use of product
formulae.

A more useful construction of a GKM algebra arises as the vertex
operator algebra of physical string states in compactifications of string
theory based on Lorentzian lattices, for example in toroidal compactifications
of all string coordinates including time
\refs{\golatt,\fgz,\moorei,\gebert}.  In our application the
metric is the Narain metric rather than the spacetime metric but similar
mathematical considerations apply.

We now discuss two specific examples
related to the product representation of threshold corrections obtained
in the previous section.

\subsec{$\Pi^{1,1}$}
We will first consider a GKM algebra with root lattice $\Pi^{1,1}$. Lattice
vectors are labelled by a pair of integers $(m,n)$ and the inner
product of two lattice vectors is $(m,n) \cdot (m',n') = -(m'n+n'm)$.
There is a single real simple root  of length squared two
which can be chosen to be
$r_{-1} = (1,-1)$.

The Weyl group is generated by reflections in the
real simple roots and so here is just the $Z_2$ transformation
acting as
\eqn\weylact{ \sigma_{-1}(r) = r - (r \cdot r_{-1} ) r_{-1} }
and thus takes $(m,n) \to (n,m) $.

A lattice Weyl vector is defined to be a lattice vector $\rho$ which
satisfies \nikulin
\eqn\lattweyl{\rho \cdot r = -r^2/2}
for any simple real root. Applying this to $r=(1,-1)$ shows that lattice
Weyl vectors must have the form $\rho = (m,m+1)$ for some integer $m$.

We now compare this to the Weyl vector $\rho$ appearing in \wvlvc\
\eqn\prodweyl{ \rho = c(-1) (0,1) - {c(0) \over 24} (1,1) }
where by the conditions given
$F(q) = \sum_{n=-1}^\infty c(n) q^n$ is a modular function (that is weight
zero).
Thus $F(q) = (j(q) -744) + c(0)$. Taking $c(-1)=1$ and using in addition the
condition that $24|c(0)$ we see that \prodweyl\ gives a lattice Weyl vector.

Borcherds has investigated two different GKM algebras related to the
$\Pi^{1,1}$ lattice. The first is the ``fake'' monster Lie algebra based on the
Lie algebra of physical states on the lattice $\Pi^{25,1} = \Lambda_{\rm Leech}
\oplus \Pi^{1,1}$. This algebra is just the algebra of physical string
states in a covariant theory with the transverse degrees of freedom
of the string compactified on the Leech lattice and the remaining space
and time dimensions on the $\Pi^{1,1}$ lattice.
The second theory is the monster Lie algebra based
roughly speaking on the Lie algebra of physical states associated with
the tensor product of the the $\Pi^{1,1}$ system with the monster vertex
algebra
of FLM based on a $Z_2$ orbifold of the Leech lattice \flm.  In this
case $c(0)=0$ since all the massless transverse states are
projected out and the Weyl vector is
 $\rho = (0,1)$.  In \firstint\ we have a product over roots
$r= (-\ell,-k)$ satisfying
the conditions $k>0$, $\ell \in Z$ or $k=0$, $\ell >0$.
Since
$c(0)=0$
 the only roots of the algebra (as compared to
the lattice) appearing in this product
are those with $k>0$, $\ell \in Z$ (and
with $\ell \ge -1$).

We will now show that
the condition $r>0$
encountered in the threshold corrections
is  the positive root condition for
a GKM algebra with Weyl vector $\rho = (0,1)$. According to a theorem
in \borchalg\ if $r$ is a positive root  then $r^2 \le - 2 \rho \cdot r $ with
equality
if and only if $r$ is simple.  With $r=(-\ell,-k)$ and $\rho= (0,1)$ as above
this
gives the inequality
\eqn\rootpcon{-2\ell k \le -2\ell }
which is satisfied for the roots $r=(-\ell, -k)$ with $k>0$, $\ell \in Z$ and
furthermore
this shows that the simple roots are the real root $(1,-1)$ and the
imaginary roots $(-\ell, -1)$ for $\ell >0$.

Using the facts
that the unique real simple root has multiplicity one and that there
are no pairwise orthogonal simple roots the denominator formula
\dengkm\ gives
\eqn\borjprod{ p^{-1} \prod_{\ell \in Z, k>0} (1- p^k q^\ell) ^{ {\rm
mult}(-\ell,-k) }
= \sum_{\ell \ge -1, n \ne 0} {\rm mult}(-\ell,-1)(p^\ell - q^\ell ) }
with $p=e^{-(0,1)}$ and $q=e^{-(1,0)}$.
As was shown by Borcherds \borchi\  the
undetermined root multiplicities can be determined indirectly based  on
the denominator formula \jprod. For Borcherds this
root multiplicity arises in the construction of the monster Lie algebra.
This leads to root multiplicities of ${\rm mult}(-\ell,-k) = c(k \ell) $ in
\borjprod\ where the coefficients $c(n)$ appear in the expansion of
the modular $j$ function with constant term set equal to zero,
$j(q) - 744 = \sum c(n) q^n $.
Here we obtain similar products but the root multiplicity is
now interpreted as the ($Z_2$ graded) multiplicity of BPS states.
The formula \borjprod\ is precisely that appearing in \firstint\
up to the $c_3(0)$ term after substituting the characters
$p=e^{2 \pi i T}$ and $q=e^{2 \pi i U } $.

For the compactification constructed previously we have $c(0)= -984$ which
leads to the Weyl vector $\rho = (41,42)$. This does not seem to lead to
a simple positive root condition for a GKM algebra. However the $\Pi^{1,1}$
case is rather exceptional in that the coefficient $c(0)$ is not determined
by the conditions on $F(q)$.  There is therefore some ambiguity in the
precise interpretation of the product formula we have obtained related to
the treatment of the $c(0)$ terms. If we explicitly evaluate these terms
in the product then the remaining product is as described above a product
over the positive roots of the monster Lie algebra. This ambiguity does
not arise for lattices $\Pi^{s+1,1}$ with $s>0$. We now turn to such an
example.

\subsec{$\Pi^{9,1}$}

In this subsection we show that the condition
$r>0$ appearing in the threshold corrections is
identical to the positive root condition for $E_{10}$.
Recall that the condition $r>0$ for
 $r = (\vec b;, -\ell,-k)$
with $s=8$ means:

\eqna\conds
$$\eqalignno{ k \ell - \half \vec b^2 & \geq -1 \quad {\rm and,} &\conds a\cr
k & >0 \qquad {\rm or,}\qquad  &\conds b\cr
k=0,\quad  \ell & > 0   \qquad {\rm or,}\qquad  &\conds c\cr
k=\ell=0,\quad  \vec b & > 0 \qquad .  &\conds d\cr}$$
where we have taken into account that
the multiplicities
$c(-r^2/2)$ vanish for $r^2 > 2$ in \conds{a}.

On the other hand, a choice of simple roots for
the $E_{10}$ lattice is given by taking:
\eqn\simprot{r_i = (\vec b ;0,0) = (\vec \alpha^{(i)};0,0) \qquad i=1,8}
where $\vec \alpha^{(i)}$ is a set of simple roots of $E_8$,
together with the extra root of $E_9$,
\eqn\enineroot{r_0 = ( -  \vec \theta; -1, 0) ,}
 ($\vec \theta $ is the highest root of
$E_8$), and
\eqn\etenroot{r_{-1} = (\vec 0 ; 1,-1). }

Given a set of simple roots $r_\mu $ the positive roots of
$E_{10}$ are simply
\eqn\posrsum{r=a_{-1} r_{-1} + a_0 r_0 + a_i r_i }
with positive $a_\mu$.
We claim this set coincides with the set $r>0$ above.

Several cases need to be checked. The only
one that causes any difficulty is the case
$k>0,\ell >0$. To check this suppose
\eqn\rti{
k \ell \geq  \half \vec b^2 -1 .
}
Let
\eqn\rtii{
a_i \vec \alpha_i = \vec b + (k + \ell) \vec \theta
}
We then want to show that $a_i>0$.

Proof: Let $\lambda_i$ be the fundamental
weights dual to $\alpha_i$. Then
\eqn\dummyk{
a_i = \vec \lambda_i \cdot \vec b + (k+\ell)
 \vec \lambda_i \cdot \vec \theta .
}
Now, by the Schwarz inequality and \rti\ we have
\eqn\dummyl{
\vert \vec \lambda_i \cdot \vec b\vert
\leq \sqrt{2k \ell + 2} \vert \lambda_i \vert
}
so
\eqn\dummym{
a_i  \geq  -  \sqrt{2k \ell + 2} \vert \lambda_i \vert
+ (k+\ell) \vec \lambda_i \cdot \vec \theta
}
Now $\vec \lambda_i \cdot \vec \theta>0$, and for
$k>0, \ell >0$ we have:
\eqn\dummyn{
(k+\ell) \vec \lambda_i \cdot \vec \theta
\geq
 \sqrt{2k \ell + 2} \vert \lambda_i \vert
}
if we satisfy
\eqn\dummyo{
(\vec \lambda_i \cdot \vec \theta)^2
\geq \vec \lambda_i^2
}
which is equivalent to
\eqn\finchk{ n_i^2 \ge G_{ii} }
with $G_{ii}$ the diagonal elements of the quadratic form of $E_8$
and $n_i$ the numerical marks (or Coxeter labels) of $E_8$.
This is true by inspection.

Moreover, the Weyl vector
$\rho = (- \vec \rho, -31, -30 )$ can be
identified with the $E_{10}$ lattice Weyl vector.
To show this it suffices to check
\eqn\chkwv{
\rho\cdot r_\mu = - \half r_{\mu}^2 \qquad \mu  = -1,0,1,\dots 8
}
As a nontrivial check on the above statements one
can prove that the vectors which satisfy the
inequality:
\eqn\brineq{
(r,r) \leq -2 (\rho,r)
}
for all positive roots of a GKM algebra
mentioned above are precisely the
vectors satisfying \conds{a,b,c,d}\ above.

We thus conclude that the product formula we obtain through fundamental
domain integrals which in turn determine the threshold corrections
can be written in terms of a product over the positive roots of $E_{10}$.
Furthermore, the product formula determines a (graded) multiplicity
associated to the roots of $E_{10}$.  The precise relation between these
multiplicities and the root multiplicities of the $E_{10}$ hyperbolic Kac-Moody
algebra is not evident to us.

\newsec{Some Limiting Cases}

Further insight into the meaning of the formula for the
prepotential can be gleaned from examining some
limiting cases.

\subsec{$M_{\rm string}\rightarrow \infty$}

In this limit we expect to  recover the standard
formulae of global N=2 supersymmetry. On the
moduli space $\CN^{10,2}$ we consider a
generic point $y_0=(\vec 0; y_+,y_-)$ on the
embedded $\CN^{2,2}$ submanifold. In
the limit $M_{\rm string}\rightarrow \infty$ a fiber of
the normal bundle of this submanifold is
identified with the vectormultiplet moduli space
of the global $E_8$ theory. Working near
$\CH^{1,1} \subset \CH^{9,1}$  we let
\eqn\mstri{
y  = y_0 + ( \delta \vec y; 0,0)
}
and in terms of dimensionful fields we let
\eqn\limcase{
\delta \vec y =  {\vec a \over  M_{\rm string} }
}
In the global limit the constraint
$2 \ypt \ymt > (\delta \vec y_2)^2$ is
trivially satisfied for fixed $\vec a$ and hence, after
modding out by the duality group we have
a coordinate $[\vec a]\in \liet(E_8)\otimes \IC/W$,
as expected.

Restoring string units, using \solution, and
the $N=2$ nonrenormalization theorems,
 the prepotential to all
orders of perturbation theory is given by:
\eqn\fullprep{
\CF = M_{\rm string}^2\biggl[
- S y^2  -
  { 1 \over  (2 \pi)^4 }
  \sum_{r>0} c_1 (-r^2/2 )  Li_3( e^{ 2 \pi  i r\cdot y} ) +\CA(y)
\biggr]
}
where $\CA(y)$ is a cubic polynomial.
Now, in taking the $M_{\rm string}\rightarrow \infty$
limit (with $\vec a$ held fixed) we may extract the
global prepotential from
\eqn\limprep{
\CF \rightarrow M_{\rm string}^2 S y_0^2 + i \CF^{\rm global}
+ \CO({\vert\vec a\vert  \over  M_{\rm string} } )
}
In order to take the limit we note that
the positive roots $r= (\vec r; -\ell, - k) >0$ with $k\not=0$ or
$\ell \not= 0$ come in pairs with $\pm \vec r$. Using this
and the limiting formula
\eqn\zslii{
Li_3 (1-x)  \rightarrow -\half x^2 \log x + \CO(x^3 \log x)
}
for the $k= \ell =0$ roots
we obtain a sum only over the positive roots of $E_8$:
\eqn\rigideff{
i \CF^{\rm global}  = \tilde S \vec a^2
- {1 \over  8 \pi^2 } \sum_{\vec \alpha >0}
(\vec \alpha \cdot \vec a)^2 \log\bigl[{2 \pi i \vec \alpha \cdot \vec a
\over  M_{\rm string} } \bigr]
}
in agreement with the standard one-loop prepotential
of the global theory, where, for any group $G$ we have
 the one-loop expressions \refs{\argf, \klti}
\eqn\zslvi{
\eqalign{
\CF & = {i \over  4 \pi} \sum_{\vec \alpha>0} (\alpha\cdot A)^2
\log{(\alpha\cdot A)^2 \over  \Lambda^2 } \cr
K & = 2 \Re\biggl[ i \CF_i \overline{A^i} \biggr] \cr
& = -{1 \over  \pi} \Re\biggl[
\sum_{\vec \alpha>0} \vert \alpha\cdot A \vert^2
( \log{(\alpha\cdot A)^2 \over  \Lambda^2 } + 1)\biggr] \cr}
}

In  \rigideff\
the renormalized ``classical'' coupling is
\eqn\clsscpl{\eqalign{
\tilde S = & S + {1 \over  8 \pi^2}
 \sum_{r>0} c_1 (-r^2/2 ) d(\vec r^2) \cr
& \bigl[
e^{ 2 \pi i r\cdot y_0} Li_2( e^{ 2 \pi i r\cdot y_0} ) +
e^{ 4 \pi i r\cdot y_0}  Li_1( e^{ 2 \pi i r\cdot y_0}) \bigr]  \cr }
}
the quantity $d(n)$ is defined by
\eqn\dndfx{
\sum_{\vec r^2 = n} ( \vec r \cdot \vec a)^2 =
d(n) \vec a^2 .
}
Finally, the cubic terms in $\CA(y)$ vanish. The
ambiguous quadratic terms with imaginary coefficients
correspond to a change in scale of $M_{\rm string}$.

\subsec{$M_{\rm string}\rightarrow 0$}

One could also try taking the opposite,
$M_{\rm string}\rightarrow 0$ limit. This raises many
new conceptual questions
\foot{and perhaps should not be attempted without
the full nonperturbative answer}
but, through the analogy with
spontaneously broken gauge theory, might
be expected to reveal fundamental underlying
symmetries in string theory \refs{\gross, \moorei, \mooreii}.

We must now introduce normalized fields for the other
two moduli
\eqn\toonew{
y_\pm = {a_\pm \over  M_{\rm string} }
}
In the low energy theory
the fields $a_\pm$ may be viewed as
Higgs fields in the Cartan subalgebra of the
 enhanced $SU(2) \times SU(2) $
or $SU(3)$ gauge symmetry.
Thus $M_{\rm string}  \rightarrow 0$ holding
$a$ fixed is equivalent
to $y \rightarrow \infty$.
More fundamentally
$a_\pm$  may be viewed
as geometrical data on $T^2$.
Consider the two-torus with vanishing $B$
field and a diagonal metric $G_{11}= R_1^2$ and $G_{22} = R_2^2$.
Then $T = 2 i R_1 R_2$ and $U= i R_2 / R_1$.
Thus the $y \rightarrow \infty$ limit is equivalent to
a decompactification to {\it five } dimensions. In this
limit the dominant terms in $\CF$ are the cubic
terms in $\CA(y)$ \foot{This cubic polynomial
is probably related to the very special geometry of
$D=5,N=2$ supergravity \dwvp}.

By the automorphic properties
of $\CF$ the limit $y \rightarrow \infty$
 is equivalent to a limit with
$y \rightarrow 0$. Taking a formal
$y \rightarrow 0$ limit leads us to
the following suggestive formulae.
We formally take the  $y \rightarrow 0$ limit
term-by-term  using \zslii\ and
\eqn\zpsli{\CP(x)  \rightarrow \pi x\bar x \log x +
\CO(x^3 \log x).}
Applying this to the formulae for  the prepotential
\fullprep\ we get:
\eqn\limcse{
\eqalign{
f^{\rm perturbative }  & \rightarrow - S (a)^2
- { 1 \over  8\pi^2}
  \sum_{r>0} c(-r^2/2)(r\cdot a)^2
\log[{2 \pi i r \cdot  a
\over  M_{\rm string} } ]\cr}
}
and the Kahler potential becomes, in the $y\rightarrow 0$
limit:
\eqn\lmkhlr{
K  \rightarrow  -\log\Biggl[ (\Im S)(-\Im a)^2
-  {1 \over  16 \pi^2} \Re\Biggl[ \sum_{r>0} c(-r^2/2)
\vert r\cdot a \vert^2 \log[
{2 \pi i r \cdot  a
\over  M_{\rm string} }  ]^2  \Biggr] \Biggr]
}
If we furthermore take a special weak-coupling
limit with $ ( \Im S ) (- \Im a)^2 \to \infty $ then
we recover the form of the global answer \zslvi\ but with
the sum running over the positive roots of $E_{10}$.

The limits \limcse\lmkhlr\ are formal since the sums
do not converge. Nevertheless
comparison with the formulae for the
global case suggests that the
weak coupling 5-dimensional theory
may be viewed, from the four-dimensional
theory as an $N=2$ gauge theory
for a generalized Kac-Moody algebra
- most of which is spontaneously broken-
with the full, infinite, spectrum of
BPS states playing the role of a
tower of  gauge bosons.
{}From the above formulae we see
that the  true root multiplicities are given by
the coefficients $c(k\ell-\half \vec b^2)$.

\newsec{An algebra associated to BPS states }

The results of sections 4-8 suggest
 that there is a GKM Lie algebra associated
to the BPS states in
D=4,N=2 heterotic compactifications.  In this section we will construct
 a GKM Lie super-algebra with even elements associated
to BPS vectormultiplets and the odd elements associated to
BPS hypermultiplets.
For the specific compactifications discussed previously this algebra has
root lattice
$\Pi^{1,1}$ or $\Pi^{9,1}$ and has root multiplicities
related to those of the BPS states.

There are two obvious puzzles in trying to construct an algebra
of BPS states.
First,
if we try to define an algebra by looking at the OPE of BPS vertex
operators then it is clear that in general the OPE will contain operators
for non-BPS states. In particular we will find operators with right-moving
conformal dimension $\bar h> 1/2$.
Second,  the physical BPS states are connected to
a lattice of the form $\Pi^{s+2,2}$ while the products involved in
threshold corrections involve only the lattice $\Pi^{s+1,1}$.   We will see
that the resolution to
the first puzzle also provides the resolution to the second.

Consider the first point. In the Neveu-Schwarz sector
a
vertex operator which creates a BPS state is of the form:
\eqn\genop{
V(z, \zb) = e^{i p_L X_L(z) + i p_R \tilde X_R(\zb)} \Phi_i(z, \zb)
}
where $ X_R$ is the bosonic component of the free superfield
in the internal  $c=3$, $N=2$ theory,  $\Phi$ is an operator of conformal
dimensions
$(h, \bar h=1/2) $ and
\eqn\dmsn{
\half p_L^2 + h - 1 = \half p_R^2 \qquad .
}
Note that  $\Phi$ is chiral or anti-chiral primary with respect  to the
the total right-moving $N=2$ SCA with $U(1)$ charge
$\pm 1$.
\foot{In the examples we consider explicitly  $X_L$ is either
the left-moving partner of $X_R$ and $(p_L, p_R) \in \Pi^{2,2} $
or we extend $X_L$ to include the lattice of the unbroken $E_8$ and
$(p_L, p_R) \in \Pi^{10,2} $. In fact our considerations are more
general and apply to any theory which has dimension one currents
on the left, whether or not they arise from a free field
construction, but for simplicity we will restrict ourselves
to that case in what follows. We will also ignore cocycle factors
in the vertex operators which are necessary to obtain a consistent
operator algebra. }

Using the decomposition \vhcont\ we can
assign to each BPS vectormultiplet
two distinguished vertex operators whose
right-moving chiral primary field in the full
internal $N=2$ theory is just the unit operator.
In the lightcone gauge this
vertex operator looks like:
\eqn\genvm{
V_{vm}^A(z) = e^{i p_L X_L(z) + i p_R \tilde X_R(\zb)} \Phi^A(z) \tilde
\psi^\mu(\zb)
}
where the index $A$ runs over an infinite range and $\tilde \psi^\mu (\bar z) $
is the right-moving transverse
spacetime fermion field, i.e., $\mu=2,3$.  Now in the OPE of two such
operators with opposite $p_R$ we will find terms of the form
$\prod \partial^n_{\bar z} X_R $.
These operators create states with right-moving oscillators in the $c=3$
theory excited, that is states which are not BPS saturated.
At this point we exploit the fact that the $c=3$ system is a
free system.
In \genvm\ we can unambiguously
drop the right-moving spacetime part
of the vertex operator and change the
right-moving free fields $X_R(\bar z)$
to left-moving free fields which we will denote by $X_R(z)$ while
at the same time retaining the Narain metric on $(p_L, p_R)$.
Technically, we have defined a mapping from the BPS
states in  the original light-cone gauge CFT
of the heterotic string:
\foot{Superscripts denote left and right conformal
charges.}
\eqn\oldcft{
\CC^{2,0}_{\rm transverse} \otimes \CC^{0,3}_{\rm transverse}
\otimes \CC^{0,3}_{\rm free\  N=(0,2)}
\otimes
\CC^{22,6}_{\rm \ N=(0,4)}
}
to new leftmoving vertex operators
of the form
\eqn\chngvm{ V_{vm}^A(z, \bar z)
\rightarrow \hat V_{vm}^A(z)\equiv
e^{i p_L X_L(z) + i p_R  X_R(z)} \Phi^A(z) }
which live in the CFT:
\eqn\newcft{
\CC^{2,0}_{\rm transverse}
\otimes
\CC^{22,6}_{\rm \ N=(0,4)} \otimes \CC^{2,0}_{\rm gaussian}
}
Note that the physical state condition means that
the left-moving conformal dimension of these vertex operators
is
\eqn\algvi{
h +\half p_L^2 - \half p_R^2 = 1
}
and hence these operators generate a current algebra based on the lattice
$\Pi^{s+2,2}$.
In terms of CFT correlators the structure constants
for three vectormultiplets are
\eqn\strccnsti{
\biggl \langle   \hat V_{vm}^1(1) \hat V_{vm}^2(2)
\hat V_{vm}^3(3)  \biggr\rangle \qquad .
}

We have now introduced two negative signature fields into
our system and the techniques of BRST cohomology are the
most appropriate to use when defining an algebra.
We also need a mechanism to kill two sets of oscillator states and at
the same time reduce the the $\Pi^{s+2,2}$ lattice  to a
$\Pi^{s+1,1}$ lattice. Such a mechanism is provided by
string theory with local $N_{\rm ws}=2$ world-sheet
supersymmetry,  particularly in the heterotic version  \oogvi.
In heterotic $N_{\rm ws}=2$ string theory the leftmoving
gauge algebra is a product
$Vir\sdtimes U(1)$. The gauged $U(1)$ current is
of the form $J = v \cdot \partial X$ with $v \in \Pi^{s+2,2}$
where $v$ is null. (For $\Pi^{2,2}$ and $\Pi^{10,2}$ there is a unique
such null vector up to lattice automorphism.)

In the present context we do not have local $N_{\rm ws} = 2$
world-sheet supersymmetry but we may nonetheless borrow these
ideas to define a Lie algebra based on a $\Pi^{s+1,1}$ lattice
which acts in a positive definite Hilbert space.  We let $L_n$
be the left-moving Virasoro operators constructed from the
$c=(26,6)$ CFT defined by \newcft\ and $J_n$ the modes of
the current defined in the previous paragraph. We then define the
physical subspace  of states as those annihilated by
the $L_n$ and $J_n$ for $n \ge 0$. The $J_0$ constraint
implies that physical states lie on a $\Pi^{s+1,1}$ sublattice
of the $\Pi^{s+2,2}$ lattice parametrizing BPS states and
the $L_n$ and $J_n$ constraints act to remove the negative norm
states associated with the two timelike oscillators.

More formally,  we enlarge the CFT to the BRST complex
\eqn\newcft{
\CC^{2,0}_{\rm transverse}
\otimes
\CC^{22,6}_{\rm \ N=(0,4)} \otimes \CC^{2,0}_{\rm gaussian}
\otimes \CC^{-2,0}_{\xi \eta} \otimes
\CC^{-26,0}_{\rm b,c\ } \otimes \CC^{2,0}_{x}
}
where the  $(\xi,\eta)$
system of $c=-2$ are the $U(1)$ ghosts. The last factor is
a $c=2$ pair of free bosonic fields which is needed to ensure
we are in the critical dimension.
Now we may  take the cohomology with respect to
$Q_{vir} $ and $ Q_{U(1)}$ which are defined by:
\eqn\algi{
\eqalign{
Q_{vir}  & = \oint \Biggl[ c\biggl( T_L(z) + \half (\p x)^2 
- \half (\p X_R)^2 +
\eta \p \xi \biggr)  - c \p c b \Biggr]\cr
Q_{U(1)} & = \oint \xi  v\cdot \p X  = \sum \xi_n v\cdot \alpha_{-n} \cr}
}
where $T_L(z)$ is the leftmoving stress tensor of the
$\CC^{2,0}_{\rm transverse}
\otimes
\CC^{22,6}_{\rm \ N=(0,4)}$ system.
It is important to work in the ``little Hilbert
space'' of the complex \FMS. Thus we should also take
cohomology with respect to $\oint \eta$.
One important effect of this cohomology
is on the allowed momentum. Since
\eqn\commu{
\{ \oint \eta, \oint \xi v\cdot \p X \} = \oint v\cdot \p X\qquad ,
}
momenta must satisfy
$p\cdot v =0 $ and are identified
mod $p \sim p + \lambda v$.
This removes momentum dependence in one $\Pi^{1,1}$ factor of the
lattice.  The cohomology with respect to the non-zero modes
of $J$ then kills two of the oscillators as desired.
We can now define the vectormultiplet Lie
algebra to be the algebra of physical states formed from
the cohomology classes of
vertex operators $\hat V_{vm}^A(z)$
with respect to the above
cohomologies.
By the no-ghost theorem the Virasoro cohomology
has a positive definite contravariant form and
thus the vectormultiplet Lie algebra is
a generalized Kac-Moody
Lie algebra (or Borcherds algebra) \borchi.

As we saw previously, threshold corrections depend on the difference
between vectormultiplets and hypermultiplets. This suggests that we
can extend this algebra to a $Z_2$ graded superalgebra with even elements
corresponding to BPS
vectormultiplets under the map \chngvm\ and the odd elements corresponding
to hypermultiplets. Threshold corrections
would then be interpreted in terms
of a supertrace over the states of this superalgebra.

This extension can be accomplished as follows.
{}From \vhcont\ we see that to each BPS hypermultiplet we can
assign two distinguished vertex operators whose
right-moving chiral primary field is a highest
weight state for the $\CA^{c=6}_{N=4}$ algebra and the unit operator for
the $\CA^{c=3}_{N=2}$ algebra.
\eqn\genvm{
V_{hm}^{A,i} (z,\zb) = e^{i p_L X_L(z) + i p_R \tilde X_R(\zb)}
\Phi^{A,i}(z,\zb)
}
where $A$ runs over an infinite range and $i$ over a finite
range (labelling the degeneracy of the $(\half,\half)$ representation
of the $N=4,c=6$ theory.)
We again construct a new vertex operator by  making $X_R$ left-moving
and dropping the unit operator in the $c=3$ theory:
\eqn\tmnvo{
V_{hm}^{A,i} (z,\zb)
\rightarrow \hat V_{hm}^{A,i} (z,\zb)\equiv
 e^{i p_L X_L(z) + i p_R  X_R(z)} \Phi^{A,i}(z,\zb)
}
This gives a vertex operator which is a holomorphic vertex operator
multiplied by an anti-holomorphic operator which is a primary
chiral operator in the $\CA^{c=6}_{N=4}$ algebra. To define
a closed algebra
we need to combine the left-moving
current algebra with the graded commutative associative
chiral algebra on the right.
Since the $N=2$
chiral algebra splits according to \chrlalg, the bosonized $U(1)$
current also splits:
\eqn\uonebos{
\sqrt{3} H^{tot} =  H + \sqrt{2} H^{N=4}
}
so that the $N=4$ current is bosonized
by:
\foot{In terms of the $N=2$ $U(1)$ current,
$J^{(2)} = 2 J_{+-}$.}
\eqn\uonebosi{
-2 J_{+-} = - i \sqrt{2} \p H^{N=4}
}
Since the $U(1)$ charge violation of the
$N=4$ system is $2$ the chiral ring of the $N=4$ theory
gives us a map from
$\hat V_{hm}^{A,i}  \times \hat V_{hm}^{B,j} \to V_{vm}^C $,
thus giving a $\IZ_2$ grading of the
desired type.  Explicitly, the structure constants
between two
hypermultiplets and one vectormultiplet are:
\eqn\strccnst{
\biggl \langle  \hat V_{hm}^1(1) \hat
V_{hm}^2(2) \hat V_{vm}^3(3)
e^{-i \sqrt{2} H^{N=4} }(\zb_4)\xi(z_0) \biggr\rangle
}
where $z_0,z_4$ can be inserted at any
point $\not=z_1,z_2,z_3$.

More formally, we can take a twisted $N=4$
cohomology \refs{\yoshii, \nojiri, \berkvaf } on the right
in addition to the Virasoro and $U(1)$ cohomologies
on the left.
{}From  the
(small) $N=4$ superconformal algebra, we have
four
odd supercurrents $G_{AB}$, $A, B = +,-$.
Now we consider a twisting so that
$T' = T - \p J_{+-}$.
Then $G_{+ A} $ has $h'=1$ and $G_{-A}$ has
$h'=2$. Thus, there are two BRST currents.
In defining the topological states we should take
{\it both} cohomologies
\eqn\nfourco{
\tilde Q_A = \oint G_{+A}
}
The resulting cohomology states are
chiral primary (not antichiral primary)
with respect to all embedded
N=2 algebras. On the resulting cohomology
we can combine the leftmoving
``Gerstenhaber bracket'' with the rightmoving
``Gerstenhaber product,''
\witzw\lianzuck,
that is, we can take the product:
\eqn\defone{
(V^1, V^2)(w, \wb)  \rightarrow \oint_w dz
\lim_{\zb \rightarrow\wb} b_{-1} V^1(z, \zb) \cdot V^2(w, \wb)
}
Under this product the right-moving
piece of two hypermultiplets maps to a
right-moving chiral primary with charge $+2$.
This does not correspond to a BPS
vertex operator, but we can conjugate, or, equivalently,
use the topological metric as in \strccnst\ above.

The vertex operators $\hat V $ associated to BPS states
then form a graded Lie  algebra of the generalized Kac-Moody type.
If we introduce vectormultiplets  with
a multiplicity two then the root supermultiplicities
of this algebra coincide with the degeneracies
$c(-r^2/2)$ occurring in the product formula.

\newsec{Application to $N=2$ String Duality}

There is now some evidence that $N=2$ heterotic string
theories are dual to Type IIA or IIB string theory on special
 Calabi-Yau spaces  \refs{\kv, \fhsv}.
These Calabi-Yau spaces have, roughly speaking, the structure of $K3$ surfaces
fibered over rational curves \cdfkm\klm. The results presented here
should allow for detailed tests of this duality in a broader
class of models than has been considered so far. We hope to
return to this issue in more detail elsewhere \wip\ but will make
a few preliminary remarks here.

The heterotic theory we have  considered
arises from the symmetric embedding of the spin connection
in the gauge group. In the $Z_2$ orbifold limit that we have
used for explicit computations the massless spectrum consists
of vector multiplets transforming in the adjoint representation
of $E_8 \times E_7 \times SU(2) \times U(1)^4$ with a total
of $388$ states (including the graviphoton) and hypermultiplets
transforming as
\eqn\hypecont{4 (1,1,1) + 8(1,56,1) + (1,56,2) + 32(1,1,2).}
with a total of $628$ states. Note that $388-628=-240$ which
matches the coefficient $c_1(0)$ in the $s=0$ case with unbroken
$E_8$ gauge group. In order to compare with the spectrum of
a possible dual Calabi-Yau space we can completely Higgs the
$E_7 \times SU(2) $ gauge group (which does not change
the difference between the number of vector and hypermultiplets)
and break $E_8$ to $U(1)^8$
by turning on Wilson lines. This leaves us with gauge group
$U(1)^{12}$ and $492$ gauge neutral
hypermultiplet fields \refs{\kv, \iban}.
The difference between
the number of massless vector and hypermultiplets is then
$12-492=-480$ which agrees with the coefficient $c_1(0)$ for
$s=8$.

This massless spectrum would arise in Type II string theory on
a Calabi-Yau space with $b_{11}= 11$ and $b_{21}= 491$ and hence
with Euler number $\chi = -960$. Precisely such a Calabi-Yau
family, denoted $X_{84}^{1,1,12,28,42}$
appears in the list of $K3$ fibrations given in \klm and is distinguished
by having the maximal value of $|\chi|$ known for Calabi-Yau spaces.
\foot{We are grateful to D. Morrison for some very
helpful remarks concerning these spaces.}
These Calabi-Yau's are resolutions of degree 84 hypersurfaces in
$\IP_4^{1,1,12,28,42}$. A typical defining polynomial would be:
\eqn\dgef{
x_1^{84} + x_2^{84} + x_3^7 + x_4^3 + x_5^2 =0 .
}
Following the procedure in \cdfkm\  we obtain a K3
fibration in the sense that there is a complete linear
system $\vert L \vert $
whose divisors are K3 surfaces. In the present
example we set $x_2 = \lambda x_1$ and define
$y_1 = x_1^2$ to get a family
$X_{42}^{1,6,14,21}$ of K3 surfaces realized as degree
42 polynomials in $\IP_3^{1,6,14,21}$. For example
\dgef\ gives:
\eqn\dgto{
(1 + \lambda^{84}) y_1^{42} + x_3^7 + x_4^3 + x_5^2 =0
}
The K3 family $X_{42}^{1,6,14,21}$ possesses
many beautiful and special properties, and has
arisen before in the physics literature
\martinec\aspmorr.  According to \aspmorr\
the family is self-mirror with complexified Kahler
cone $\cong \CH^{9,1}$.

The formulae we have derived for the heterotic
prepotential fit in well with the predictions of
heterotic/type II  duality. According to \kv\klm\ the
cohomology class in $H^2(X;Z)$ dual to a divisor
$L$ in the linear system defines the heterotic
coordinate $S$. The remaining generators of the
K\"ahler cone define coordinates $y\in \CH^{9,1}$
and may be identified with the K\"ahler classes
of $X_{42}^{1,6,14,21}$.  The form of the
$S$-dependence, $S y^2 $ of the classical prepotential
follows from the structure of the $K3$-fibration,
therefore, let us consider the third derivatives
with respect to the $y^A$. The
third derivative of the perturbative prepotential
following from \solution\ is
\eqn\crvcount{
{\p \over  \p y^A} {\p \over  \p y^B} {\p \over  \p y^C}
\CF^{\rm pert} =
{\tilde d^{10,2}_{ABC} \over  64 \pi^2} + {1 \over  2 \pi}
\sum_{r>0} c_1(-r^2/2) r_A r_B r_C {e^{-2 \pi r \cdot y} \over
1 - e^{-2 \pi r \cdot y}  }
}
Note this is in precisely the right form expected for
the counting of rational curves on a Calabi-Yau.
Substituting the values of $\tilde d^{10,2}_{ABC}$
from the appendix we see that the third derivatives
of $\CF$ are indeed integers if we multiply
$\CF$ by $2 \pi t $ where $t$ is an integer.
\foot{In verifying this it is important to
bear in mind the freedom to redefine
$\CF$ by \ambig. We hope to further investigate
this chamber-dependence in
the future \wip.}
Moreover, since $c_1(0) = \half \chi$
we now recognize the constant term in
\solution:
\eqn\theterm{-{2 \delta \over 384 \pi^2} =
 - { c_1(0) \zeta(3)\over 2 (2 \pi)^4 }, }
as the famous $\zeta(3)$ term
appearing in the Calabi-Yau prepotential of \cogp\  which
is related to the four-loop renormalization of the Calabi-Yau
sigma-model \gvz. Comparing with the
normalization of \cogp\ we see that the properly
normalized prepotential must be
$4 \pi \CF$.

In view of this  heterotic/type II duality
makes some curious  predictions for
algebraic geometry, namely:

The cubic intersection product on $H^{1,1}$
for the family $X_{84}^{1,1,12,28,42}$
is governed by the
structure of $E_8$ and may be computed
from
the expressions in $(A.51)$ and $(A.52)$
by differentiation.
Moreover, the  rational
curves in the family $X_{84}^{1,1,12,28,42}$
 which are orthogonal
to $c_1(L) \in H^2(X; \IZ)$  are
parametrized by the positive roots $r>0$
of $E_{10}$ and appear with multiplicity
$2 c_1(-r^2/2)$ where $c_1(n)$ are
the coefficients of the modular form
$F(q) = E_6/\eta^{24} $.

If the rational curves are not isolated
the integers must be interpreted as
integrals over the moduli space, as
is standard in topological field theory.
These integers should be related to
the numbers of rational curves on
$K3$ surfaces in $X_{42}^{1,6,14,21}$.

Remarks entirely analogous to the
above apply to the relation between
the family of $K3$ fibrations
$X_{24}^{1,1,2,8,12}$ discussed in
\kv\klm\  and the formula \yukawa.

\newsec{Concluding Speculations}

We have shown that threshold corrections in $N=2$ heterotic string
theories are determined by the spectrum of BPS states and have
provided evidence that there is a generalized Kac-Moody algebra
associated with these states which governs the form of the threshold
corrections.  Various extensions of these results should prove
very interesting.

It would be interesting to generalize the computations
here to other backgrounds and more general dependence
on the moduli. In particular, the computations done here involved
the standard embedding of the spin connection on $K3$ in the
gauge group. By repeating these calculations for different
ranks of gauge group and for different topologies of
gauge bundle it should be possible to recover
interesting product formulae associated with a
large number of hyperbolic and generalized
Kac-Moody algebras.
As examples, we expect to get an
interesting algebra associated with
$\Pi^{17,1}$ even with the symmetric
embedding, and moreover
we expect to recover the products
occurring in the work of Feingold and Frenkel
\fein. Indeed, the relevant products have
already been suggested in \mayrsti.

Other embeddings and other choices of moduli dependence will
also lead to modular integrals for congruence subgroups of
$SL(2,\IZ)$.  Given the presence (for the
standard embedding) of the Monster Lie algebra and
the $j$ function, which is the Thompson series for the identity element
of the Monster, it seems likely that these other
embeddings will involve some of the other Thompson series for
the Monster. It might be that the Monster provides a general
classification of $N=2$ heterotic string vacua.

The formula we have given for the perturbative
heterotic prepotential fits in well with the proposed
heterotic/type II string duality and therefore
admits a natural extension to a full nonperturbative
answer.  In this
paper we have shown that the special Kahler geometry of the
$N=2$ heterotic compactification is
summarized by the prepotential:
\eqn\mnrslt{
\eqalign{
\CF  = & - S (y,y) +
{1 \over  384 \pi^2}  i \tilde d^{s+2,2}_{ABC} y^A y^B y^C
- {\zeta(3) \over  2 (2 \pi)^4} c_1(0)  \cr
& -  { 1 \over  (2 \pi)^4 }
  \sum_{r>0} c_1 (-r^2/2 )  Li_3( e^{ 2 \pi i r\cdot y} )
 + \CF^{\rm nonpert}(y, e^{2 \pi i S} )\cr}
}
where $\CF^{\rm nonpert}(y, e^{2 \pi i S} )$ has
an analytic power series expansion in its second argument.
As noted in the previous section, the formula for the
prepotential as a sum of trilogarithms is
{\it dictated} by the curve-counting
formulae given a dual type II background. Thus, for
backgrounds admitting a dual pair, the full nonperturbative
prepotential will again be a sum of trilogarithms, with
the replacement
$$e^{ 2 \pi i r\cdot y} \rightarrow e^{ 2 \pi i r\cdot y + 2 \pi i n S} $$
in the argument of the trilogarithms. The sum will
run over the full positive Kahler cone of the dual
Calabi-Yau variety. In view of our results it is
natural to speculate that this sum will again be
a sum over positive roots of some interesting algebra.

The results of this paper should have
two interesting mathematical applications in
the context of heterotic/type II string duality.
First,
given an $N=2$
dual string pair the results presented here combined with those
of \bcov\ suggest that special combinations of Ray-Singer torsions
on dual Calabi-Yau spaces  should admit infinite product representations.
These considerations are undoubtedly related to the
recent work of Jorgenson and Todorov \jorgenson.
Second, interesting recent work
of Lian and Yau \lianyau, has shown that certain
mirror maps associated to one-parameter families of
algebraic K3 surfaces are related to Thompson series.
Since the heterotic string is a more natural home for
the Monster it is tempting to speculate that the
algebraic structure of BPS states together with
string duality might provide a way to understand the
observations of \lianyau.

The algebraic structures we have begun to uncover
should also shed new light on string duality. For example,
it is tempting to speculate that different dual theories are simply
different representations of an underlying algebraic structure much
as in the different realizations of affine Kac-Moody algebras.  It would
be very interesting to try to identify generalized Kac-Moody algebras
in Type II string theories on Calabi-Yau spaces arising
as $K3$ fibrations.
In addition to the applications to $N=2$ string
duality outlined in the previous section our work
might also have relations to the conjectured
duality between heterotic and Type I theories \wittdyn.
The exchange of $\bar z$ and $z$ dependence needed to
define the vertex algebra is reminiscent of world-sheet orbifolds
considered in \refs{\horava, \sgntti, \polch} and recently applied to
Type I -heterotic duality  in \vwdual.  There is possibly
an alternative point of view on our construction which might be interesting
to pursue.  In closed string theory there is a Virasoro algebra which
generates transformations on the spatial string coordinate $\sigma$
with generators $\hat L_n = L_n - \bar L_{-n} $. If we take the $\bar L_{-n}$
to be those of the free right-moving bosons and the $L_{n}$ those of the
left-moving degrees of freedom then the vertex operators we have
defined are dimension one as a result of the condition \algvi. If we
were then to complexify $\sigma$ we would obtain a holomorphic
vertex operator algebra. This suggests a complexification of the string
world-sheet, something that has also been suggested in other
contexts \wittorb.

Finally, if the GKM algebra associated to BPS states that we have
found is a gauge algebra, as it appears,
then in analogy to the structure of $N=2$
Yang-Mills theory  the full nonperturbative prepotential
should be governed by  ``monopoles''
of this algebra and the monodromies
given by an algebraic variety
whose monodromy group is the Weyl group of
this algebra.  There is a natural candidate for these
monopoles. Recall that the gauge structure became
clearest in the $M_{\rm string} \rightarrow 0$ limit,
which coincides with a decompactification to
5 dimensions. We propose that the states being
counted by the product representation of
threshold corrections are five-branes
wrapped around the remaining $K3\times S^1$
internal space.

\bigskip
\centerline{\bf Acknowledgements}\nobreak
We thank the Aspen Center for Physics where this work was begun for
providing a stimulating atmosphere.
This work was  supported in part by NSF Grant No.~PHY 91-23780 and
DOE grant DE-FG02-92ER40704.

We would like to thank S. Ferrara,
A. Losev, J. Louis, D. L\"ust, E. Martinec, D. Morrison,
H. Nakajima,
N. Nekrasov,
H. Nicolai, M. O'Loughlin, S. Shatashvili, S. Shenker,
A. Taormina, and A. Todorov for discussions. J.H owes
special thanks to Neam Harvey.

\appendix{A}{Fundamental Domain Integrals}

In this appendix we evaluate the following two integrals:
\eqn\genint{
\CI_{s+2,2}(y)\equiv
\int_{\CF} {d^2 \tau \over  \tau_2}
\Biggl[
\biggl(\sum_{p\in \Gamma^{s+2,2}}
 q^{\half p_L^2} \bar q^{\half p_R^2}\biggr)
F(q) - c(0) \Biggr]
}
and
\eqn\geninti{
\tilde \CI_{s+2,2}(y)\equiv
\int_{\CF} {d^2 \tau \over  \tau_2}
\Biggl[
\biggl(\sum_{p\in \Gamma^{s+2,2}}
 q^{\half p_L^2} \bar q^{\half p_R^2}\biggr)
F(q) (E_2-{3 \over  \pi \tau_2})-\tilde c(0)\Biggr]
}
These integrals are absolutely convergent
for $y\in \CH^{s+1,1}$ where $y$ is not on
the RQD's. They have logarithmic singularities
on the RQD's and moreover, are
$O(s+2,2;\IZ)$ invariant.

The notation is as follows:
$\CF$ is the fundamental domain for $SL(2,\IZ)$.
$F(q) $ is a  $SL(2,\IZ)$ modular function of weight $-s/2$ for
$\CI$ or of weight
$-s/2-2$ for $\tilde \CI $ which is holomorphic except at
infinity where it has a first order pole.
That is,
we assume $F$ has a Fourier expansion in terms of $q=e^{2 \pi i \tau}$:
\eqn\srs{
F(q)  =\sum_{n=-1}^\infty c(n) q^n
= c(-1) q^{-1} +c(0) + \cdots
}
and define $c(n) \equiv 0$, $n< -1$.
We also define coefficients $\tilde c(n) $ through the expansion:
\eqn\tlcfss{
F(q) E_2(q) \equiv \sum_{n=-1}^\infty \tilde c(n) q^n
}
where $E_2(q)$ is the first of the series of Eisenstein functions:
\eqn\eetwo{
E_2(q) = 1 - 24 \sum_{n = 1}^{\infty} {n q^n \over 1 - q^n}
= 1 - 24 \sum_{n = 1}^{\infty}{ \sigma_1(n) q^n } \; ,
}
\eqn\eefour{
E_4(q) = 1 + 240 \sum_{n = 1}^{\infty} {n^3 q^n \over 1 - q^n}
= 1 + 240 \sum_{n = 1}^{\infty}{ \sigma_3(n) q^n } \; ,
}
\eqn\eesix{
E_6(q) = 1 - 504 \sum_{n = 1}^{\infty} {n^5 q^n \over 1 - q^n}
= 1 - 504 \sum_{n = 1}^{\infty}{ \sigma_5(n) q^n } \; ,
}
where
\eqn\esigma{
\sigma_k(n) = \sum_{d | n} d^k
}
is the sum of the $k$th powers of the divisors of $n$.

$E_2(\tau)$ is not modular covariant since it transforms with a shift. The
combination
\eqn\covrnt{
E_2(\tau) - { 3 \over  \pi \tau_2}
}
appearing in \geninti\ on the other hand is not holomorphic but
transforms covariantly.

The sums in \genint\ and \geninti run over an
even self-dual lattice  $\Gamma^{s+2,2}(y)
\cong \Pi^{s+2,2}$, and thus
$s=8t$ with $t$ integer. As discussed in the text
we write $\Pi^{8t+2,2} \cong \Pi^{8t,0} \oplus \Pi^{2,2}$
and write lattice vectors as
\eqn\fdintvii{
(\vec b; m_+, n_-; m_0, n_0)
}
with metric
\eqn\fdintviii{
(\vec b; m_+, n_-; m_0, n_0) ^2 = \vec b^2 -2 m_+ n_- + 2m_0 n_0
}
Also, for Narain compactifications we have
\eqn\fdintix{
\eqalign{
\half (\vec p_L^2 - \vec p_R^2) & = \half \vec b^2 - m_+ n_- + m_0 n_0\cr
\half p_R^2 & = {1 \over  -2 (\Im y)^2} \biggl\vert
\vec b \cdot \vec y -  m_+ y_-  -  n_- y_+ + m_0 -
 \half n_0 y^2 \biggr\vert^2\cr}
}
Our conventions for the complex coordinates on the
homogeneous space $SO(8t+2 ,2)/SO(8t+2) \times SO(2)$
are as in sec. 2.

\subsec{Evaluation of $\CI_{s+2,2}$ }

We will follow quite closely the calculation in the
appendix
of \dkl.
The general strategy for the evaluation of \genint\ and \geninti\
is to perform first a Poisson resummation on the ``momenta''
$m_+$ and $m_0$ which leads to a sum over matrices $A$
which are general two by two matrices with integer elements.
The contributions of two matrices $A$, $A'$ related by an element $V$of \SLZ
are related by modular transformation of $\tau$ by $V$ which
allows one to sum  instead over orbits of \SLZ and integrate
over the images of the fundamental domain under the elements
$V$ that yield distinct matrices $A$ when acting on a representative
element of the orbit.

We now consider \genint.  Performing a Poisson resummation on
$m_+$, $m_0$ gives
\eqn\fdintxi{
\tau_2 Z_\Gamma=
\tau_2 \sum_{p\in \Gamma^{s+2,2}}
 q^{\half p_L^2} \bar q^{\half p_R^2}
=\sum_{A\in Mat_{2\times 2}} T_\Gamma[A] }
where
\eqn\interdum{ T_\Gamma[A]
= {-(\Im y)^2 \over  2 y_{-,2} } \sum_{\vec b\in \Gamma^{8s,0} }
  q^{\half \vec b^2} \exp{{\cal G}},
}
\eqn\fdintxij{\eqalign{ {\cal G} =&
\Biggl[ {\pi \imysq \over  2 \ymt^2 \tau_2} \vert \CA\vert^2
- 2 \pi i y_+ \det A
+ {\pi \over  \ymt} \vec b \cdot (\yv \tilde \CA - \yvb \CA) \cr
& -{\pi \over 2 \ymt} n_0 (\yv^2 \tilde \CA - (\yvb)^2 \CA)
+
{i \pi \over  \ymt^2} \imyvsq\ (n_- + n_0 y_-^*) \CA \Biggr]\cr }
}
and
\eqn\fdintxii{
\eqalign{
\CA& = \pmatrix{ 1 & y_- \cr} \pmatrix{n_-& k_1\cr n_0 & -k_2\cr}
 \pmatrix{\tau\cr 1 \cr}= \pmatrix{ 1 & y_- \cr} A  \pmatrix{\tau\cr 1 \cr}\cr
\tilde \CA& = \pmatrix{ 1 & y_-^* \cr} \pmatrix{n_-& k_1\cr n_0 & -k_2\cr}
 \pmatrix{\tau\cr 1 \cr}\cr}
}
Under modular transformations we have
\eqn\fdintxiii{
\eqalign{
A  & \to A \pmatrix{a & b \cr c& d \cr} \cr
\CA & \to {1 \over  c \tau + d} \CA\cr}
}
Unlike the simpler integral analyzed in \dkl, it is no longer
obvious that the contribution from two matrices $A$
related by the modular transformation \fdintxiii\  is given
by the modular transformation $\tau' = {a \tau + b \over c \tau + d } $.
However for $s=8t$
 it is still true as can be seen by Poisson resummation
on the $\Pi^{8t,0}$ lattice sum and using the fact that
$\Pi^{8t,0}$ is even self-dual.  Note that one could
equally well have used a Poisson resummation on
$n_-,m_0$ leading to the same expression with
$y_+$ exchanged for $y_-$.

Following \dkl\ we can split the sum on $A$ into three orbits
and correspondingly write the integrals as a sum of
three integrals:
\eqn\throrbs{
\eqalign{
\CI_{s+2,2} & = \CI_{s+2,2}^0 + \CI_{s+2,2}^{nd} + \CI_{s+2,2}^{dg} \cr
\CI_{s+2,2}^0 & =
\int_{\CF} {d^2 \tau \over  \tau_2^2} T_{\Pi}[A=0] F(q) \cr
\CI_{s+2,2}^{nd} & =
2 \int_{\CH}
{d^2 \tau \over  \tau_2^2} \sum_{0\leq j <k,p\not=0}
T_{\Pi}[A=\pmatrix{k& j\cr 0 & p\cr} ] F(q) \cr
\CI_{s+2,2}^{dg} & =
 \int_{\CS}
{d^2 \tau \over  \tau_2^2}\biggl[ \sum'_{j,p}
T_{\Pi}[A=\pmatrix{0& j\cr 0 & p\cr} ] F(q)
- \tau_2 c(0) \theta(\tau\in \CF) \biggr] \cr}
}
where $\CH$ is the upper half plane and
$\CS$ is the strip $ \{ \tau \in \CH, | \tau_1| < 1/2  \} $. These three
terms correspond to the zero orbit $A=0$, the non-degenerate orbit
with $det A \ne 0 $ and the degenerate orbit with $det A=0$ and a
particular choice of representative matrices from each orbit.

In deriving \throrbs\ we have made an important
exchange of summation and integration. Because
of the tachyon divergences $\sim q^{-1}$ which exist
before the $L_0 = \bar L_0$ projection this exchange
can be invalid. Indeed, it is clear that
if $\ypt> 2 \ymt$ then one must
use the expressions \fdintxi,\interdum\fdintxij.
On the other hand, if $\ymt> 2 \ypt$, one must use
the other Poisson summation with
$y_+ \leftrightarrow y_-$.  In fact, we know that the
expressions must change as we cross the wall
of the Weyl chamber $\ymt = \ypt$.
\foot{Hence there is a puzzle about why the exchange
of sum and integral is not valid for one Poisson summation
in the range $2> \ypt/\ymt > 1$. We have not
resolved this point.}

The $A=0$ orbit may be evaluated using  \lsw\  \foot{After correcting
a factor two error in eqn.  9.38 of \lsw.} .
\eqn\fdintxivpr{
\eqalign{
{-(\Im y)^2 \over  2 y_{-,2} }  \int_{\CF} {d^2 \tau \over  \tau_2^2}
\vartheta_{\Pi^{8t,0}} F(q)
&= {-(\Im y)^2 \over  2 y_{-,2} }  {1 \over  \pi}
[G_2(q) \vartheta_{\Pi^{8t,0}} F(q)]_{q^0} \cr}
}
For $t=1$ which will be of most relevance for our discussion
we have
\eqn\fdintxiv{
  {1 \over  \pi}
[G_2(q) \vartheta_{\Pi^{8,0}} F(q)]_{q^0}  =
 {\pi \over  3} (c(0)+216 c(-1))
}

The non-degenerate orbit is evaluated using the
representative matrix
\eqn\mmrep{A_0=\pmatrix{k & j \cr 0 & p\cr}  }
with $0\leq j <k$ and $p\not=0$.
The $\tau_1$ integral is gaussian and the sum on $j$ is then trivial.
The integral over $\tau_2$ gives a representation of the Bessel
function $K_{1/2}$ which is elementary.
The sum on $p$ yields a logarithm so that the contribution from
the non-degenerate orbit is given by
\eqn\fdintxv{
-2 \log\Biggl\vert \prod_{\vec b\in \Pi^{8t,0}}
\prod_{k>0,\ell\in \IZ} \biggl(1-e^{2 \pi i \hat{r \cdot y}  }
 \biggr)^{c(k\ell - \half \vec b^2)}
\Biggr\vert^2
}
where we have introduced a ``hatted dot product '' defined by
\eqn\notation{
\hat{r \cdot y}  \equiv \Re \biggl[
(\vec b \cdot \vec y
+ \ell y_- + k y_+)\biggr]
+ i \biggl\vert  \Im \biggl[
(\vec b \cdot \vec y
+ \ell y_- + k y_+)\biggr] \biggr\vert
}
when $k>0$, and, when $k=0$:
\eqn\notationp{
\eqalign{
\hat{r \cdot y}  & \equiv
r\cdot y - N(r,y) y_-\cr
N(r,y) & = \sgn(\vec b) \biggl[ \sgn(\vec b)  {\vec b \cdot \Im \vec y \over
y_{-,2} } \biggr] \cr}
}
where
$[ \cdot]$ is the greatest integer function.

To evaluate the contribution from the degenerate orbit
we take
\eqn\degreps{A_0=\pmatrix{0 & j \cr 0 & p\cr}  }
with $j,p$ not both zero. After evaluating
the $\tau_1$ integral  the sum on $j$ may
be performed using a Sommerfeld-Watson transformation:
\eqn\sommwat{
\eqalign{
\sum_{j=-\infty}^\infty { e^{i \theta j} \over  (j+B)^2 + C^2}
& =  {\pi \over  C} e^{- i \theta (B-iC) } {1 \over  1- e^{- 2 \pi i (B-iC)} }
\cr
& +
 {\pi \over  C} e^{- i \theta (B+iC) }
{e^{2 \pi i (B+iC)} \over  1- e^{2 \pi i (B+iC)} } \cr
& C>0, \qquad 0 \leq \theta \leq 2 \pi\cr}
}
A special case is
\eqn\spck{
\sum_{j=1}^\infty { \cos \theta j \over  j^2}
= {\theta (\theta-2 \pi) \over  4} + {\pi^2 \over  6}
}
for $0\leq \theta \leq 2 \pi$.

To write the final answer it is useful to introduce the
polylogarithm functions:
\eqn\newfns{
\eqalign{
Li_1(x) & = \sum_{j=1}^\infty  {x^j \over  j} = - \log(1-x) \cr
Li_2(x) & = \sum_{j=1}^\infty {x^j \over  j^2}   = \int_0^1 {dt \over  t}
{1\over  (1-xt) }\cr
Li_3(x)  & = \sum_{j=1}^\infty {x^j \over  j^3}  = - \int_0^1  {dt \over  t}
\int_0^1  {ds \over  s}  \log(1-xts)\cr } }

The general answer for the integral
\genint\ is then
\eqn\ansi{
\eqalign{
\CI_{8t+2,2}(y) & =  {-(\Im y)^2 \over  2 y_{-,2} }  {\pi \over  3}
[E_2(q) \vartheta_{\Pi^{8t,0}} F(q)]_{q^0} \cr
& + c(0) \Biggl( - \log [-(\Im y)^2] + {\pi \over  3} y_{-,2}  - {\cal K }
\Biggr) \cr
& + c(-1) {2 y_{-,2}\over  \pi} \sum_{\vec b^2=2} \Re\Biggl[
Li_2\bigl( e^{2 \pi i {\vec b \cdot \Im \vec y \over  y_{-,2} } } \bigr)
\Biggr]\cr
& +
4\Re \Biggl\{
\sum_{r>0} \biggl[
 c(kl - \half \vec b^2)  Li_1(e^{2 \pi i \hat{r\cdot y}}) \biggr]\Biggr\} \cr}
}
Here $r>0$ means:

1. $k>0$ or,

2. $k=0$, $\ell>0$ or,

3. $k=\ell=0$, $\vec b>0$.

The first line of \ansi\
comes from $A=0$. The second and
third lines come from the degenerate orbit,
and the last line comes from the nondegenerate
and degenerate orbits. The constant ${\cal K}$ is
\eqn\clkdef{
{\cal K}  = \log[{ 4 \pi \over  \sqrt{27} } e^{1-\gamma_E} ] }
where $\gamma_E$ is the Euler-Mascheroni constant.

We now consider two special cases.

\noindent $\underline{s=0}$:
In this case it is customary to denote
$y_+ = T, y_- = U$.
The general formula \ansi\ then gives
\eqn\tutup{
\eqalign{
\CI_{2,2}(T,U) & = c(0)[- \log [ 2 T_2 U_2] -{\cal K} ]
 + c(0) {\pi \over  3} (T_2 + U_2) \cr
& -2 \log\biggl\vert e^{-2 \pi i T c(-1) } \prod_{r>0}
\Biggl(1-e^{2 \pi i (k T+ \ell U) }
\biggr)^{c(k\ell)}
\Biggr\vert^2 \cr}
}
Here
$r>0$ means
$k>0, \ell\in \IZ$, or
$k=0, \ell>0$.  We have also assumed that $T_2 > U_2$. In this case
one may replace $\hat{r\cdot y}$ with $r\cdot y$ since they  only differ
for $k=1, \ell=-1$ and in this case $\hat{r\cdot y} = (T_1-U_1) + i \vert T_2 -
U_2 \vert$
which equals $r \cdot y $ for $T_2 > U_2$. For $U_2 > T_2$  one should
interchange $U$ and $T$ in \tutup.

The last two terms in  \tutup\ may be written as
\eqn\fdintv{
-2 \log \vert j(T) - j(U) \vert^2 - 2 \log
\vert \eta(T) \eta(U) \vert^{2c(0)}
}
using the product formula \jprod.

\noindent $\underline{s=8}$:
In this case we have, without loss of generality,
\eqn\seight{
F(q) = { E_4^2 \over  \eta^{24}}
}
The finite sums over
simple roots of $\Pi^{8,0}$
can be simplified.  In \ansi\ we have a term
\eqn\spck{\Re\biggl[Li_2(e^{i\theta})\biggr]=
\sum_{j=1}^\infty {\cos \theta j \over  j^2}
= {\theta (\theta-2 \pi) \over  4} + {\pi^2 \over  6}
\qquad \quad  0\leq \theta \leq 2 \pi
}

In order to apply this to the third line of \ansi\
we need to be careful to take care of the
range of the angle
${\vec b \cdot \Im \vec y \over  \ymt}$.
Since, by assumption, $y$ is not at an enhanced symmetry
point
this quantity is not integral.
If we further take
$y$ to be in the fundamental Weyl chamber \intrvl\
 we get:
\eqn\pzero{
2 c(-1) {2 \ymt \over  \pi}
\sum_{\vec b^2 =2, \vec b>0} \Biggl[{ \pi^2 \over  3}
- 2 \pi^2 {\vec b\cdot \Im \vec y \over  \ymt} +
2 \pi^2  ({\vec b \cdot \Im \vec y \over  \ymt} )^2
\Biggr]
}

We can evaluate the sum over $E_8$ root vectors.  The Weyl
vector of $E_8$ is
$  \vec \rho  \equiv \half \sum_{\vec b^2 =2, \vec b>0} \vec b  $
and the quadratic sum can be evaluated \patera\  to give
\eqn\rts{
\sum_{\vec b^2 =2} (\vec b\cdot \vec v)^2  = 60 \vec v^2 }

To summarize, under the assumption \intrvl\ we get the product:
\eqn\fdintxx{
\eqalign{
\CI_{10,2}(y)
& =- 2 \log \bigl\vert
\Phi(y) \bigr\vert^2
+ c(0) \Biggl( - \log [-(\Im y)^2]   - {\cal K} \Biggr)
 \cr
\Phi(y) & = e^{2 \pi i \rho\cdot y}
\prod_{r>0 }
 \biggl(1-e^{2 \pi i \hat{r \cdot y}  }
 \biggr)^{c(k\ell - \half \vec b^2)}
\cr
\rho & = -(\vec \rho; 31,30) \cr}
}

In general the integral $\CI_{s+2,2}(y)$ has a Weyl
vector. This means that the cubic terms cancel, as
can be derived using theorems 6.2 and 10.3 of
Borcherds \borchii.
Moreover,
the Weyl vector $\rho$
appearing in the product formulae of
\borchii\  can be extracted from the  terms linear in $y$.
Abstractly we have - for all $t \geq 1$:
\eqn\wylv{
\ypt {\pi \over  3} [E_2 \vartheta_{\Pi^{8t,0}} F(\tau)]_{q^0}
+
\ymt {\pi \over  3} [ \vartheta_{\Pi^{8t,0}} F(\tau)]_{q^0}
-4 \pi  \sum_{\vec b>0} c(-\vec b^2/2) \vec b \cdot \Im \vec y .
}
This agrees exactly with the expression in theorem 10.4 of \borchii.

\subsec{Evaluation of $\tilde \CI$}

We now turn to  an evaluation of \geninti.
We again decompose the sum over $A$ into a sum
over orbits of $SL(2,\IZ)$.  The $A=0$ orbit  is done
using  \lsw
\eqn\lswfrm{
\int_{\CF} {d^2 \tau \over  \tau_2^2}
(\hat G_2 (\tau))^k F(q)
={1 \over  \pi(k+1)} \biggl[(G_2 (\tau))^{k+1} F(q)\biggr]_{q^0}
}
The evaluation of the non-degenerate orbit proceeds as before except that
the $\tau_2$ integral now gives a representation of $K_{3/2}$ and in the
sum over $p$ one makes the replacement
\eqn\repb{
{1 \over  \vert p \vert} \rightarrow
{ \biggl\vert  \Im \biggl[
(\vec b \cdot \vec y
+ \ell y_- + k y_+)\biggr] \biggr\vert \over  p^2}
+ {1 \over  2 \pi  \vert p \vert^3}
}
Thus, instead of logarithms we get $Li_2(x)$ and
$Li_3(x)$. The evaluation of the degenerate orbit also
proceeds as before except that the sum on $j$ now involves
\eqn\nwswt{
\sum_{j=-\infty}^\infty { e^{i \theta j} \over  ((j+B)^2 + C^2)^2}
 = -{1 \over  2 C} {\p \over  \p C}
\sum_{j=-\infty}^\infty { e^{i \theta j} \over  ((j+B)^2 + C^2)} }
which thus reduces to \sommwat.  The result is
\eqn\ansii{
\eqalign{
\tilde \CI_{8t+2,2}(y)  = & {-(\Im y)^2 \over  2 y_{-,2} }  {\pi \over  6}
[E_2^2(q) \vartheta_{\Pi^{8t,0}} F(q)]_{q^0}
 + \tilde c(0) ( - \log [-(\Im y)^2]
+ {\pi \over  3} y_{-,2}  - {\cal K}  ) \cr
+ &
c(0) {2 \pi \over 15} {\ymt^3 \over (\Im y)^2}
+ {6 \over  \pi^2} {c(0) \over  (\Im y)^2}  \zeta(3) \cr
 +  & \sum_{\vec b^2=2}   \Re\Biggl[
\tilde c(-1) {2 y_{-,2}\over  \pi}
Li_2\bigl( e^{2 \pi i {\vec b \cdot \Im \vec y \over  y_{-,2} } }
\bigr)+ c(-1) {12 \ymt^3 \over \pi^3 (\Im y)^2}
Li_4\bigl( e^{2 \pi i {\vec b \cdot \Im \vec y \over  y_{-,2} } } \bigr)
\Biggr]
\cr
 + &
4\Re \Biggl\{
\sum_{r>0} \biggl[
 \tilde c(kl - \half \vec b^2)
Li_1(e^{2 \pi i \hat{r\cdot y}})
+ {6 \over  \pi (\Im y)^2} c(kl - \half \vec b^2) \CP(\hat{r\cdot y}) \biggr]
\Biggr\}
 \cr}
}
where we have also introduced the function
\eqn\funkyfun{ \CP(x)  = \Im(x) Li_2(e^{2 \pi i x} ) +
 {1 \over  2 \pi} Li_3(e^{2 \pi i x} )  .}
In fact,  this integral can be written in
the form:
\eqn\thrshxi{
\eqalign{
\tilde \CI_{8t+2,2}(y)  =  & 4\Re \Biggl\{
\sum_{r>0} \biggl[
 \tilde c(kl - \half \vec b^2)
Li_1(e^{2 \pi i \hat{r\cdot y}})
+ {6 \over  \pi (\Im y)^2} c(kl - \half \vec b^2) \CP(\hat{r\cdot y}) \biggr]
\Biggr\}
 \cr
& +  \tilde c(0) \Biggl( - \log [-y_2)^2]
  - {\cal K} \Biggr)
 + {1 \over  (y_2)^2} [ \tilde d^{8t+2,2}_{ABC} y_2^A y_2^B y_2^C
+ \delta ]  \cr}
}
where the constant is
\eqn\cnst{
\delta = {6 \over  \pi^2} c(0)  \zeta(3)
}
and $\tilde d$ is a real symmetric tensor, which depends on
the Weyl chamber.

We now consider the two special cases $s=0$ and $s=8$.
For the case $s=0$ without loss of generality we have
\eqn\fedef{
F = {E_4 E_6 \over \eta^{24}}
}
With $y_+ = T, y_- = U$ as before we have
\eqn\ttutup{
\eqalign{
\tilde \CI_{2,2}(y)  = &
4\Re \Biggl\{
\sum_{r>0} \biggl[
 \tilde c(kl )
Li_1(e^{2 \pi i \hat{r\cdot y}})
- {3 \over  \pi T_2 U_2 } c(kl ) {\cal P}(\hat{r\cdot y}) \biggr]
\Biggr\}
 \cr
- { \delta \over  2 T_2 U_2}  & - 264 [-\log[2 \ypt\ymt] - {\cal K} ]\cr
& - 48 \pi T_2 - 88 \pi U_2
+ 16 \pi {U_2^2 \over  T_2} \qquad T_2 > U_2 \cr
& - 48 \pi U_2 - 88 \pi T_2
+ 16 \pi {T_2^2 \over  U_2} \qquad U_2 > T_2 \cr}
}
As discussed above $\hat{r\cdot y}$ is only different from
$r\cdot y$ for $k=1, \ell=-1$ and in this case it
is $(T_1-U_1) + i \vert T_2 - U_2 \vert$.

When $s=8$,  we have
\eqn\axdgy{
F = { E_6 \over  \eta^{24}}
}
To calculate $\tilde \CI$ we need
\eqn\spckii{
\Re\biggl[Li_4(e^{i\theta})\biggr]=
\sum_{j=1}^\infty { \cos \theta j \over  j^4} =
{\pi^4 \over 90}  - {1\over 48} \theta^2 (2\pi - \theta)^2
}
valid for $0\leq \theta\leq 2\pi$.
We also need the $E_8$ root sum
\eqn\eatep{
\eqalign{
\sum_{\vec b^2=2,\vec b>0} (\vec b \cdot \vec v)^4 & = 18 (\vec v^2)^2\cr
}
}
where the quartic sum can be evaluated using formulae in \patera.
 Again specializing
to the Weyl chamber \intrvl\ we can write the answer
as in \thrshxi\  with
 the symmetric tensor $\tilde d$ is determined by:
\eqn\cubctns{
\eqalign{
\tilde d^{10,2}_{ABC} y_2^A y_2^B y_2^C
 = & - 8 \pi \biggl[ (\vec \rho \cdot \vec y_2 + 41 \ymt + 42
\ypt)  (y_2)^2 \cr
- 2 \sum_{\vec b^2=2,\vec b>0} (\vec b \cdot  \vec y_2)^3
& + 60 \ymt^2 \ypt + 72 \ypt^2 \ymt + 4 (\ymt)^3 \biggr]\cr}
}
Finally, we need an explicit formula for the cubic
sum over $E_8$ roots. This is given by
\eqn\cubate{
\eqalign{
\sum_{\vec b^2=2,\vec b>0} (\vec b \cdot  \vec v)^3
= &
6\,{{v(1)}^2}\,v(2)
+ 2\,{{v(2)}^3} + 6\,{{v(1)}^2}\,v(3) +
  6\,{{v(2)}^2}\,v(3)  \cr
& + 4\,{{v(3)}^3} +
 6\,{{v(1)}^2}\,v(4) +  6\,{{v(2)}^2}\,v(4) + 6\,{{v(3)}^2}\,v(4) \cr
& +
 6\,{{v(4)}^3}
 +
  6\,{{v(1)}^2}\,v(5) + 6\,{{v(2)}^2}\,v(5) +
 6\,{{v(3)}^2}\,v(5) \cr
& +
6\,{{v(4)}^2}\,v(5) + 8\,{{v(5)}^3} + 6\,{{v(1)}^2}\,v(6) +
  6\,{{v(2)}^2}\,v(6)  \cr
& +
6\,{{v(3)}^2}\,v(6) +
  6\,{{v(4)}^2}\,v(6) +6\,{{v(5)}^2}\,v(6) + 10\,{{v(6)}^3} \cr
& +
6\,{{v(1)}^2}\,v(7) +
  6\,{{v(2)}^2}\,v(7) + 6\,{{v(3)}^2}\,v(7) +
  6\,{{v(4)}^2}\,v(7) \cr
& +
6\,{{v(5)}^2}\,v(7) + 6\,{{v(6)}^2}\,v(7) + 12\,{{v(7)}^3} +
  30\,{{v(1)}^2}\,v(8)  \cr
& + 30\,{{v(2)}^2}\,v(8) +
 30\,{{v(3)}^2}\,v(8) + 30\,{{v(4)}^2}\,v(8) + 30\,{{v(5)}^2}\,v(8) \cr
& + 30\,{{v(6)}^2}\,v(8) +
  30\,{{v(7)}^2}\,v(8) + 22\,{{v(8)}^3} \cr}
}
where $v(i)$ is the $i^{th}$ component of $v$ in the usual
Cartesian basis for $E_8$ roots.

\listrefs
\bye